\documentclass[aps,pre,showpacs,amsfonts,amssymb,amsmath,floatfix,twocolumn,
floats,nobalancelastpage]{revtex4}

\usepackage{graphicx}

\begin{document}

\title{
Roughening of the $(1+1)$ interfaces in two-component surface growth\\
with an admixture of random deposition 
}

\author{A. Kolakowska}
\author{M. A. Novotny}
\author{P. S. Verma}
\affiliation{Department of Physics and Astronomy, and the ERC
Center for Computational Sciences,
P.O. Box 5167, Mississippi State, MS 39762-5167}

\date{\today}

\begin{abstract}
We simulate competitive two-component growth on a one dimensional substrate of $L$ sites. 
One component 
is a Poisson-type deposition that generates Kardar-Parisi-Zhang 
(KPZ) correlations. The other is 
random deposition (RD). We derive the universal scaling function of the interface 
width for this model and show that the RD admixture acts as a dilatation mechanism 
to the fundamental time and height scales, but leaves the KPZ correlations intact. 
This observation is generalized to other growth models. It is shown that the 
flat-substrate initial condition is responsible for the existence of an early 
non-scaling phase in the interface evolution. The length of this initial phase 
is a non-universal parameter, but its presence is universal. 
We introduce a method to measure the length of this initial 
non-scaling phase. In application to parallel and distributed computations, the important 
consequence of the derived scaling is the existence of the upper bound 
for the desynchronization in a conservative update algorithm for  
parallel discrete-event simulations. It is shown that such algorithms are 
generally scalable in a ring communication topology.
\end{abstract}


\maketitle

\section{INTRODUCTION \label{intro}}

In (1+1) dimensions, the roughness of a surface that grows on a one dimensional 
substrate of length $L$ can be expressed by the interface width $w(t)$ at time $t$
\begin{equation}
\label{width}
\langle w^2 (t) \rangle =\left\langle \frac{1}{L} \sum_{k=1}^L 
\left( h_k (t) - \bar{h}(t) \right) ^2 \right\rangle,
\end{equation}
where $h_k(t)$ is the height of the column at site $k$ and $\bar{h}(t)$ is the average 
height. The angular brackets denote the average over many configurations 
and the bar over a symbol denotes the average over $L$ sites. 
The self-affined roughness of the interface manifests itself by the presence of 
the Family-Vicsek (FV) scaling \cite{FV85}:
\begin{equation}
\label{Family1}
w^2(t) = L^{2 \alpha} f \left( \frac{t}{L^z}\right),
\end{equation}
where the scaling function $f(y)$ describes two regimes of the width evolution:
\begin{equation}
\label{Family2}
f(y) \sim \left\{ \begin{array} {r@{\quad , \quad}l}
y^{2 \alpha /z} & y \ll 1 \\
\mathrm{const.} & y \gg 1.  \end{array} \right. 
\end{equation}
The dynamic exponent $z$ gives the evolution of the lateral correlation 
length $\xi (t) \sim t^{1/z}$. When $\xi (t)$ exceeds the system size $L$ the 
width saturates. At saturation, for $t \gg t_{\times}$, the width scales as 
$w \sim L^{\alpha}$, where $\alpha$ is the roughness 
exponent. The growth phase is the initial phase for $t \ll t_{\times}$ before the 
cross-over time $t_{\times} \sim L^z$ to saturation. The growth phase is characterized 
by the single growth exponent $\beta = \alpha /z$. The roughness, growth and  
dynamic exponents are universal. Their values depend only on the underlying mechanism 
that generates the correlations and scaling. 

A simple continuum model of nonequilibrium growth that leads to the scaling of 
a noise-driven interface is provided by the Kardar-Parisi-Zhang (KPZ) 
equation \cite{KPZ86}. In the co-moving frame, the KPZ equation is
\begin{equation}
\label{kpz}
h_t = \nu \, h_{xx} + \frac{\lambda}{2} \, h_x^2 + \zeta \,  ,
\end{equation}
where $h=h(x,t)$ is the height field (subscripts denote partial derivatives). 
Coefficients $\nu$ and $\lambda$ give the strength of the linear damping and the 
coupling with nonlinear growth, respectively. The uncorrelated Gaussian noise 
$\zeta (x,t)$ has zero mean and covariance
\begin{equation}
\label{noise}
\langle \zeta (x,t) \zeta (x',t') \rangle = D \, \delta (x-x') \delta (t-t') ,
\end{equation}
where $D$ is the noise strength. A renormalization group analysis \cite{KPZ86,BS95} 
can provide a connection between the stochastic growth equation and scaling exponents. 
The KPZ universality class, governed by dynamics given by Eq.~(\ref{kpz}), is 
characterized by $\alpha = 1/2$ and $\beta = 1/3$. A characteristic signature of 
the KPZ scaling is the exponent identity $\alpha + z =2$, valid in all dimensions. 
When $\lambda =0$ in Eq.~(\ref{kpz}), the growth is governed by the linear 
Edwards-Wilkinson (EW) equation \cite{EW82}. The EW universality class is characterized 
by $\alpha = 1/2$ and $\beta = 1/4$. Using scaling arguments, it can be shown 
that in (1+1) dimension the EW exponent identity is $2 \alpha + 1 = z$ \cite{Kru97}. 
When $\lambda =0$ and $\nu =0$ in Eq.~(\ref{kpz}), the growth belongs to the RD 
universality class, characterized by $\beta = 1/2$ and the lack of saturation. 
The RD interface is not self-affined.

The theory behind kinetic roughening and the origins of scale invariance 
are well understood \cite{BS95,YWL93,FV91,Kru97,Kar96,Odor04}, but there are numerous 
instances of growth processes that neither follow one power law nor exhibit a 
clear-cut universality as it is expressed by the FV scaling. One group 
of examples is the anomalous roughening in epitaxial growth models \cite{Kru97,DL+96,DDK96}, 
fractures \cite{LS98,MSL+98} and in models with subdiffusive behavior or quenched 
disorder \cite{LR96}. These systems exhibit different dynamic scaling on local and 
global scales, characterized by different values of roughness exponents. 
The super-rough dynamics of tumor growth \cite{BP+98} is the first experimental 
observation of anomalous scaling in (1+1) dimension. Another issue is the clear 
experimental observation of the KPZ universality and the role of quenched noise in 
the asymptotic KPZ scaling \cite{Kru97,Kar96,CC01}. For one dimensional KPZ 
growth, by applying a weak noise canonical phase-space method, it has been shown 
recently that the KPZ dynamic exponent is associated with the soliton dispersion 
law \cite{Fog02}. However, at saturation all KPZ correlations are exactly the same 
as would result from the linear EW equation \cite{Kru97,Fog02,FOR+94}. 
The fact that the EW equation is sort of ``embedded" in the 
KPZ equation may give rise to ambiguous values of scaling exponents for growth 
mechanisms (or models) that interpolate between the weak and the strong nonlinear 
coupling regimes. Even when the nonlinear coupling is strong the discrete models 
require sufficiently large length and time scales to show clear KPZ scaling 
\cite{AF90,AF91,HF91,NT92,BE01}. Likewise, large scales are essential in simulation 
studies of roughening in the two-component growth models that combine one 
process governed by linear EW dynamics with another process governed by nonlinear 
KPZ dynamics \cite{PJ91}. Recently, Chame and Reis \cite{CR02} simulated in (1+1) 
dimension a mixed growth where particles aggregated either by ballistic deposition 
(with probability $p$) or by random deposition with surface relaxation 
(with probability $1-p$). They show that for small $p$ and sufficiently large $L$, 
the interface width has three well-defined evolution stages. The first stage, 
for early times, is the EW growth. The second stage, for intermediate times before 
saturation, is the KPZ growth. The third stage is the saturation.

Two distinct growth phases were also 
observed in  experiments with interfacial roughening in Hele-Shaw flows 
\cite{HS+01,SOH02,SOH03}, in simulations of electrophoretic deposition of polymer 
chains \cite{FP98,BPF00} and in numerical studies of one dimensional restricted 
solid-on-solid models with two growth components, each of which 
being a mechanism ruled by dynamics that belongs to a distinct 
universality class  \cite{Gry03,KPS98,KSP98,WC93,EC00,Reis02,HMA01,HA01}. The 
latter examples suggest that the two-phase growth may be an effect of mixing the 
universalities. In support of this hypothesis comes a two-component model considered 
by da Silva and Moreira \cite{DSM01}, where the Kim-Kosterlitz \cite{KK89} deposition 
occurs with probability $p$ and ballistic deposition occurs with probability $(1-p)$. 
Both of these growth mechanisms are governed by the KPZ dynamics, except that their 
corresponding continuum equations differ in the sign of the coupling $\lambda$. 
In this case, large scale simulations in (1+1) dimension produce only one growth 
phase with the growth exponent $\beta$ parametrized by $p$ \cite{DSM01}.

The purpose of roughening studies presented in this article is
to uncover the dynamics of desynchronization in a conservative
parallel discrete-event simulations (PDES). The PDES
are a technical tool to uncover the
dynamics of information-driven complex stochastic systems. Their wide range of
applications in contemporary sciences and technology \cite{Fuj00} has
made them an active area of research
in recent years. Parallel and distributed simulation systems constitute
a complex system of their own, whose properties can be uncovered with the
well-established tools of statistical physics.

In PDES physical processes are mapped to logical processes (assigned to processors)
that manage state-updates of assigned physical subsystems. The main challenge arises
because logical processes are not synchronized by a global clock. Consequently, to
preserve causality in PDES the algorithms should incorporate the so-called local
causality constraint \cite{CM79,Fuj90}. Depending on the way the local causality
constraint is implemented, there are two broadly defined classes of update protocols
\cite{Fuj00}: conservative algorithms \cite{CM79,Mis86,Lub87,Lub88} and optimistic
algorithms \cite{Jef85,DR90,Ste93}. Important efficiency considerations of these
algorithms involve the utilization of the parallel processing environment
(as measured by a fraction of processors working simultaneously at a time)
and the memory per processor required by state savings. The latter is closely related
to the statistical spread in the processors' local times, i.e., to desynchronization. Recent
applications of conservative PDES to modeling complex physics systems include ballistic
deposition \cite{LPR96}, Ising spins \cite{KNR99} and dynamic phase transition
\cite{KWRN01,KRN02}. In an application to simulating
Ising spins, an onset of self-organized critical
behavior in optimistic PDES has been recently reported \cite{SOS01,OSS02}.

Since the introduction by Korniss {\it et al.} \cite{KTN+00} an idea of utilizing
nonequilibrium surface growth methods \cite{BS95,YWL93,FV91} in evaluating
the scalability of algorithms for conservative PDES, there have been a number of
advances using such efforts. The main concept behind this idea is the virtual time
horizon (VTH) of the algorithm. The VTH is a (1+1) dimensional nonequilibrium surface.
Its time evolution can be simulated by applying a deposition rule that is defined by
a parallel-update protocol based on the algorithm. Several properties of the
algorithm can be deduced from analyzing its corresponding simulated VTH. One of
them is the utilization \cite{KNR03,KN04}.
Another one is the desynchronization of processors in the system as the PDES evolves
in time. The width of the simulated VTH provides a measure of this desynchronization.
It has been demonstrated that an asymptotic lack
of synchronization \cite{KNK+02} can be avoided in new generations of algorithms
\cite{KNK03,KNG+03}. The focus of past studies
was on the worst-case performance scenario when each parallel process
consisted of only one computational object (or computational volume).
Despite being informative (e.g., providing the evidence of a lower bound on the
utilization)
such studies are of little practical value since taking full advantage
of parallelization implies many computational objects per processor.
Past simulation studies have demonstrated that when the load per processor
is increased the utilization increases significantly
\cite{KNK03,KNR03,KN04,NKN03}, and a mean-field-like method to estimate
the utilization in this general case has been developed \cite{KNR03,KN04}. The missing
element thus far has been a detailed investigation of the dynamic scaling properties
of the simulated VTH for the general case of many computational objects per processor.
This is the main theme of this article.

We simulate three growth mechanisms for the VTH that correspond
to three implementations of a conservative update algorithm in a ring
communication topology, where each processor communicates only with its immediate
neighbors. The models are variations of Poisson-random depositions.
Two of these rules are realizations of the worst-case scenarios and
present a situation when the system of $L$ processors can be mapped onto a
closed chain of $L$ spins. The third model corresponds to the general case
when each processor carries $N$ computational volumes and it
combines the second model with random deposition (RD). In this
work, we do not attempt to obtain exact values of scaling exponents
that characterize these growth processes. Our primary interests are in the
scaling functions for the interface width
and in the universal properties of the VTH interfaces.

Simulation studies of the VTH interface in the worst-case scenario showed  
that for very large $L$ this interface belongs to the KPZ universality class 
\cite{KTN+00}. However, for small $L$ or for early times, before the KPZ growth is 
attained, the width does not scale. This suggests the strong sensitivity of the 
evolution to the initial condition. When the model is generalized to 
accommodate many computational volumes per processor, the evolution of the 
VTH width changes. Now there are two distinct phases in the growth regime. 
The early phase evolves in the RD fashion and the later phase 
has signatures of the KPZ scaling. In this work we investigate the above issues 
in large-scale simulations. Unless stated otherwise, configurational 
averages were obtained over an ensemble of 800 independent simulations. The VTH 
models and definitions are explained in Sec.~\ref{models}. 
Simulations are initiated from a flat substrate and carried 
on up to $10^7$ time steps, well beyond cross-over times to the steady state for the 
considered substrate sizes. In Sec.~\ref{flat} we analyze the interface evolution 
for random depositions at local surface minima (i.e., the worst-case scenario) 
and show that the initial lack of scaling is an artifact of the flat-substrate 
initial condition. To identify nonuniversal features in the evolution, in addition 
to Poisson-random depositions we also consider both Gaussian and uniform-random depositions. 
In the steady-state time averages we omit the index $t$ in the notation, 
e.g., $\langle w^2 \rangle$ denotes the saturated surface width. 
In Sec.~\ref{scaling} we perform the analysis of the interfaces 
generated by two simultaneously acting growth mechanisms, one of which being 
RD and the other generating the KPZ correlations, and we obtain a 
universal scaling function for this type of VTH interfaces. Results obtained 
in this section show that the RD admixture  
elongates the principal height and time scales,  
leaving the KPZ correlations intact. In Sec.~\ref{velocity} we derive a general 
relation between the VTH interface velocity and the utilization in conservative 
update processes. In Sec.~\ref{roughening} we generalize findings of 
Sec.~\ref{scaling} to two-component models that mix RD  
with a deposition that either classifies within the KPZ or within the EW 
universality class. In particular, we show that the RD admixture that happens 
with probability $(1-p)$ gives rise to a $p$-dependent affine component in the scaling. 
Section~\ref{roughness} contains the discussion of finite-size effects observed 
in scaling of the VTH interfaces. An example of false 
scaling in Sec.~\ref{false} is provided to illustrate the importance of the 
relaxation from the 
flat-interface initial condition in the scaling considerations. Applications to 
scaling and scalability of conservative PDES algorithms are discussed 
in Sec.~\ref{application}. Conclusions are summarized in Sec.~\ref{summary}.

\section{SIMULATION MODELS \label{models}}

In simulations a system of $L$ processors is represented as a set of equally 
spaced lattice points $k$, $k=1, 2,...,L$. Each processor performs a number of 
operations and enters a communication phase to exchange information with its 
immediate neighbors. This communication phase, called an update attempt, takes 
no time in our simulations. In this sense we simulate an ideal system of processors 
(the relation to PDES is discussed in Sec.~\ref{application}). An update attempt is 
assigned an integer index $t$ that has the meaning of a wall-clock time 
(in arbitrary units).

The local virtual time $h_k(t)$ at the $k$th processor site represents the cumulative 
local time of all operations on the $k$th processor from the beginning at 
$t=0$ to time $t$. These local processor times are not synchronized by a global clock. 
The ring communication topology among processors 
is mapped onto a lattice arrangement with periodic 
boundary conditions, $h_{L+k}(t)=h_k(t)$.
The set of local virtual times $h_k(t)$ forms the VTH at $t$.  
The growth of the VTH is simulated 
by a deposition rule, where local height increments $\eta_k(t)$ are sampled 
from the Poisson distribution of unit mean. The form of the deposition rule depends 
on the processor load, as explained below.

A general principle that governs the conservative update protocol requires a processor 
to idle if at the update attempt $t$ the local causality constraint may be violated. 
This happens when at $t$ the $k$th processor does not receive 
the information from its neighboring processor (or processors) 
if such information is required to proceed in its computation. 
This corresponds to a situation when the local virtual time $h_k(t)$ 
of the $k$th processor is ahead of either one of the local virtual 
times $h_{k-1}(t)$ or $h_{k+1}(t)$ of its left and right neighbors, 
respectively. In this unsuccessful update attempt the local virtual 
time $h_k(t)$ is not incremented, i.e., the $k$th processor waits: 
$h_k(t+1) = h_k(t)$. In another case, for example, when at $t$ the 
$k$th processor does not need information from its neighbors it performs an update 
regardless of the relation between its local virtual time and the local virtual times on 
neighboring processors.

One example of computations that follow the above model is a dynamic Monte Carlo 
simulation for Ising spins. In a parallel environment, a spin lattice is spatially 
distributed among $L$ processors in such a way that each processor carries an equal 
load of one contiguous sublattice that consists of $N$ spin sites (i.e., each processor has a 
load of $N$ volumes). Some of these $N$ spin-lattice sites belong to border slices, 
i.e., at least one of their immediate neighbors resides on the sublattice of a 
neighboring processor. Processors perform concurrent spin-flip operations 
(i.e., increment their local virtual times) as long as a randomly selected 
spin-site is not a border site. If a border spin-site is selected, to perform 
a state update that is consistent with and faithful to the underlying physical 
spin dynamics, a processor needs to know the current spin-state of the corresponding 
border slice of its neighbor. If this information is not available at the $t$ update 
attempt (because the neighbor's local time is behind), by the conservative update 
rule the processor waits until this information becomes available, i.e., until the 
neighbor's local virtual time catches up with or passes its own local virtual time.

The least favorable parallelization is when each processor carries one computational 
volume, $N=1$. Computationally, this system can be identified with a closed spin 
chain where each processor carries one spin-site. At each update attempt each 
processor must compare its local virtual time with the local times on both of its neighbors.

The second least favorable arrangement is when processors have a computational volume $N=2$. 
As before, the system can be mapped onto a closed spin chain where each processor 
carries two spin-sites, each of which is a border site. At each update attempt every processor 
must compare its local time with the local time of one of its neighbors.

In general, when $N \ge 3$, at update attempt $t$, the comparison of the local virtual 
times between neighbors is required only if the randomly selected volume site is from 
a border slice. In all cases, we start the simulation from a flat substrate 
at $t=0$, $h_k(0)=0$.

At every successful update attempt, the simulated local virtual time at the $k$th  
site is incremented for the next update attempt: $h_k (t+1) = h_k (t) + \eta_k (t)$, 
where $\eta_k(t)= - \ln (r_{kt})$, and $r_{kt} \in (0;1]$ is a uniform random deviate. 
The three cases described above are realized in simulations by the following three 
deposition rules.\\
{\bf Rule 1 $(N=1)$:} The update attempt at $t$ is 
successful iff
\begin{equation}
\label{rule1}
h_k (t) \le \min \left\{ h_{k-1} (t), h_{k+1} (t) \right\}.
\end{equation}
{\bf Rule 2 $(N=2)$:} At any site $k$ where the update attempt 
was successful at $(t-1)$, at $t$  we first randomly select a neighbor 
(left or right). This is equivalent to selecting either the left or 
the right border slice on the $k$th processor. The update 
attempt is successful iff
\begin{equation}
\label{rule2}
h_k (t) \le h_n (t),
\end{equation}
where $n$ is the randomly selected neighbor ($n=k-1$ for the left, 
$n=k+1$ for the right). At any site $k$ where the update attempt 
was not successful at $(t-1)$, 
at $t$ we keep the last $n$ value.\\
{\bf Rule 3 $(N \ge 3)$:} At any site $k$ where the update attempt 
was successful at $(t-1)$, at $t$  we first randomly select any of the 
$N$ volume sites (indexed by $n_k$) assigned to a processor. 
The selected site can be either from the border sites (either $n_k=1$ 
or $n_k=N$) or from the interior. The attempt is successful if the selected site is the 
interior site. When the border site is selected, the attempt is successful if 
condition~(\ref{rule2}) is satisfied. As in Rule $2$, at any site $k$ where the 
update attempt was not successful 
at $(t-1)$, at $t$ we keep the last $n_k$ value.

Rule $3$ is essentially different from rules $1$ and $2$ in that it is a mixture 
of rule $2$ and RD. At each $t$, depending on the selected $n_k$, 
the local update (deposition) at site $k$ either follows rule $2$ that requires 
checking with a neighbor or follows RD that just simply deposits a random positive 
real number $\eta_k(t)$. The probability $(1-p)$ that the rule $3$ takes the form of RD is 
parametrized by $N$. 
At each $t$, the complementary probability $p$ that rule $3$ 
takes the form of rule $2$ can be obtained by a direct count of 
lattice sites that have the assigned value either $n_k=1$ or $n_k=N$, and 
subsequently taking the configurational mean of this count. The mean 
density of these sites can be interpreted as the probability $p(N)$ 
that at $t$ a randomly selected site followed rule $2$. 
In steady-state simulations (defined in Sec.~\ref{flat}) this density 
does not depend on $t$ and is found to be approximately $p=\sqrt{2/N}$. 
This is because in simulating conservative PDES the random selection 
of volume site $n_k$ is not performed at every $t$ for all lattice sites $k$. 
If at some $t$ the update condition~(\ref{rule2}) is not satisfied, the selection of 
$n_k$ must be postponed until some later time $t'$ when 
condition~(\ref{rule2}) is satisfied. Explicitly, in the spirit of conservative 
PDES, if relation~(\ref{rule2}) does not hold then the old $n_k$ is kept 
for as many update steps as required until it finally holds at later $t'$. 
Note, if the draw of $n_k$ were performed at each $t$ for all $k$ the 
probability of selecting a border site would have been $2/N$, i.e., smaller than $p$.

We define the utilization $\langle u (t) \rangle$ as the configurational average 
of the fraction of sites that performed an update at $t$. When $N=1$, 
$\langle u (t) \rangle$ is simply the mean density of local minima of the interface. 
When $N \ge 2$, $\langle u (t) \rangle$ is the mean density of update sites. 
The velocity $v(t)$ of the interface is defined as $\langle d \bar{h}(t)/dt \rangle$.

One distinction between the deposition models studied here and other restricted 
solid-on-solid models is that in the former the deposited random height increment 
$\eta$ is a positive real number that can take on any value from an assigned real 
interval, while the latter usually consider integer height increments. Although 
in the context of applications to PDES this is the Poisson distribution with mean 
$\mu _P =1$ and unit variance that models the waiting times, represented here by 
$\eta$, we also consider two other alternative depositions. One of them is  
uniform deposition, where $\eta$ is sampled from the  uniform distribution with 
mean $\mu _U =1/2$ and variance $1/12$, restricted to the interval $(0;1]$. 
The other one is a Gaussian deposition, 
where $\eta$ is sampled from the Gaussian of unit variance and zero mean, 
restricted to the positive semi-axis. For this Gaussian, after nolmalization, 
the mean is $\mu _G = \sqrt{2/\pi}$ and the variance is $(1-2/\pi)$. We find that varying the 
deposition type does not change the universality class of the model. The purpose 
of introducing this variation is to better identify nonuniversal features in 
the initial evolution of the VTH interface.

\begin{figure}[bp]
\includegraphics[width=6.6cm]{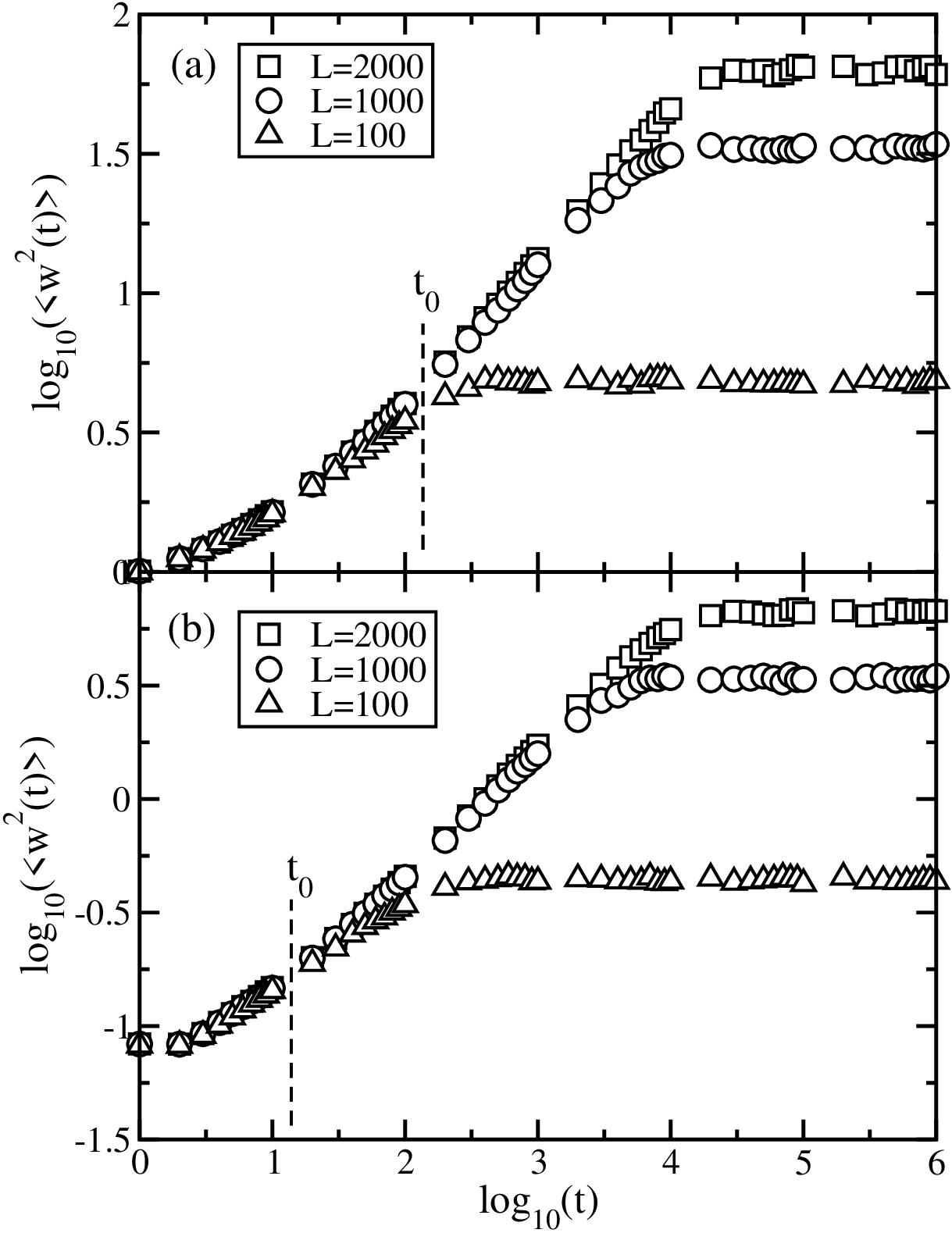}
\caption{\label{pre04f01}
Time evolution of the simulated VTH width when $N=1$ for various sizes $L$.
(a) Poisson deposition. (b) Uniform deposition.
The initial phase for $t < t_0$ does not scale.
}
\end{figure}

\begin{figure}[bp]
\includegraphics[width=6.6cm]{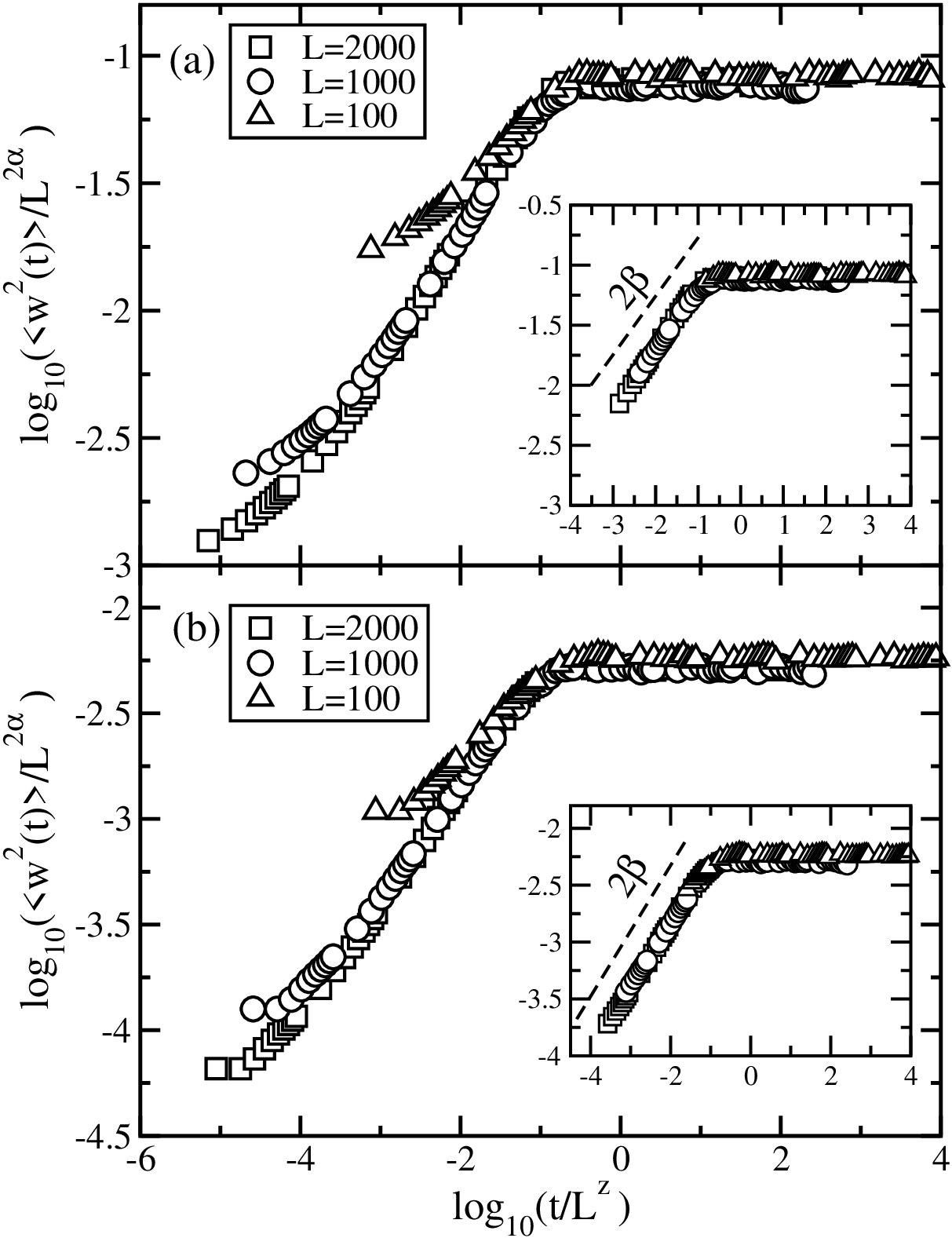}
\caption{\label{pre04f02}
Scaling of the VTH width when $N=1$. (a) Poisson deposition.
(b) Uniform deposition. The inserts show the data collapse for $t>t_0$.
The slope $2\beta$ of the growth phase is consistent with $\alpha + z =2$
and $\alpha = \beta z$.
}
\end{figure}

\section{FLAT-SUBSTRATE CONDITION \label{flat}}

Rule $1$ $(N=1)$, is a realization of random deposition at local
surface minima. As demonstrated in simulations with Poisson depositions 
\cite{KTN+00}, these interfaces belong to the KPZ universality class when
the system size $L$ is sufficiently large. Using standard finite-size scaling
techniques \cite{BS95} for $L$ of the order $10^3$-$10^5$,
the scaling exponents were determined numerically
to be $\beta = 0.326 \pm 0.005$ and $\alpha = 0.49 \pm 0.01$ \cite{KTN+00}.
A peculiarity of this scaling, present for all $L$, is the existence of the
initial phase that does not scale, as illustrated in Fig.~\ref{pre04f01}{\it a}.
Another feature is the lack of scaling for small $L$, approximately $L<100$. 
In this section we analyze
the interface evolution for processes that obey rule $1$ with Poissonian, uniform and
Gaussian depositions, and show that the above lack of scaling is an artifact of the
flat-substrate initial condition.

Figure~\ref{pre04f01} presents typical evolutions of the interface width
for moderate to large $L$
for both Poissonian and uniform depositions. A similar behavior is also seen 
for Gaussian depositions. A common
feature is the existence of an initial growth phase, $0 \le t <t_0$, where the widths
do not scale. For $t > t_0$ the widths obey the FV scaling,
Eq.~(\ref{Family1}-\ref{Family2}), with KPZ exponents.
Figure~\ref{pre04f02} presents the FV scaling for Poissonian
depositions (with $2\alpha = 0.88$) and for uniform depositions (with $2\alpha = 0.94$), 
both with $\alpha + z = 2$, 
when the scaling transformation is applied for all $t \ge 0$. A similar picture of the
data collapse is obtained for Gaussian depositions (with $2\alpha = 0.92$).
The whisker-like structures in the growth part, clearly observed in
Fig.~\ref{pre04f02}, demonstrate the absence of scaling for $0 \le t <t_0$.
They vanish when the scaling is restricted to times $ t \ge t_0$
(the inserts to Fig.~\ref{pre04f02}) and full data collapse is
achieved for these later  times. This initial transition phase
is not a finite-size effect since $t_0$ does not depend on $L$. For all $L$ there
is one common $t_0$ that depends only on the deposition type. The largest $t_0$
is observed for the deposition with the largest variance of the
random height increments $\eta$. In our examples, the smallest $t_0$ is for the
uniform depositions (variance $1/12$) and the largest $t_0$ is for Poissonian
depositions (variance $1$). In Gaussian depositions (variance $(1-2/\pi)$)
the initial $t_0$ falls between these two values. Thus, while the
scaling shows that the mechanism of generating correlations
(i.e., rule $1$) has KPZ dynamics, the length
$t_0$ of this initial relaxation period to KPZ scaling is not universal.

\begin{figure}[tp]
\includegraphics[width=6.6cm]{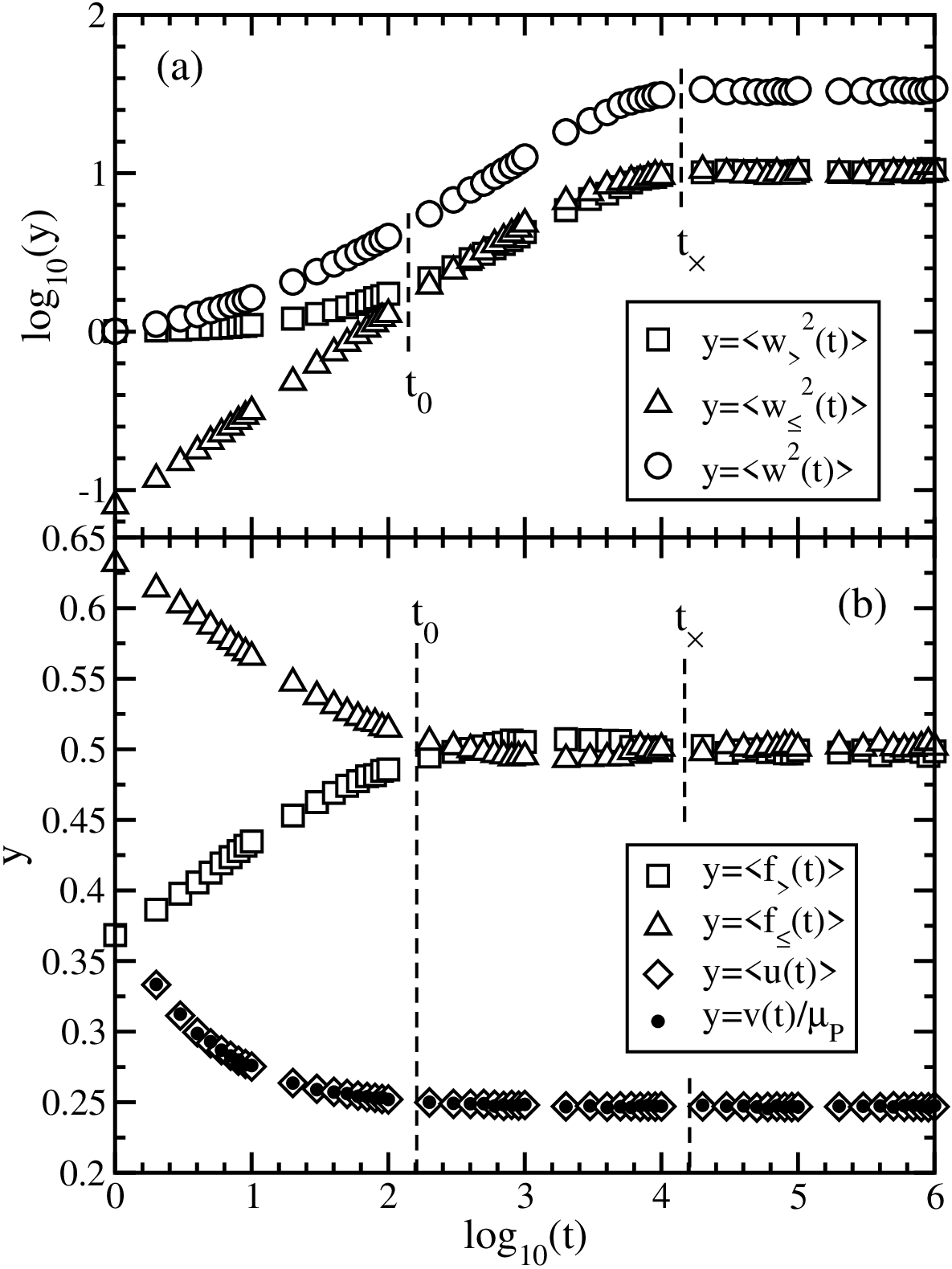}
\caption{\label{pre04f03}
Time evolutions for $N=1$ with Poisson deposition $(L=1000)$.
(a) The widths, $w^2 = f_{\le}w_{\le}^2 + f_{>}w_{>}^2$.
(b) The interface velocity $v(t)$ and characteristic densities:
the utilization $\langle u(t) \rangle$ ($\langle u(0) \rangle=1$, not shown),
the simplex coefficients $\langle  f_{>}(t) \rangle$ and $\langle  f_{\le}(t) \rangle$.
The time $t_0$ marks the transition to the steady state
(the KPZ growth), and $t_{\times}$ is the cross-over time to saturation.
Here, $\mu _P =1$.
}
\end{figure}

\begin{figure}[tp]
\includegraphics[width=6.6cm]{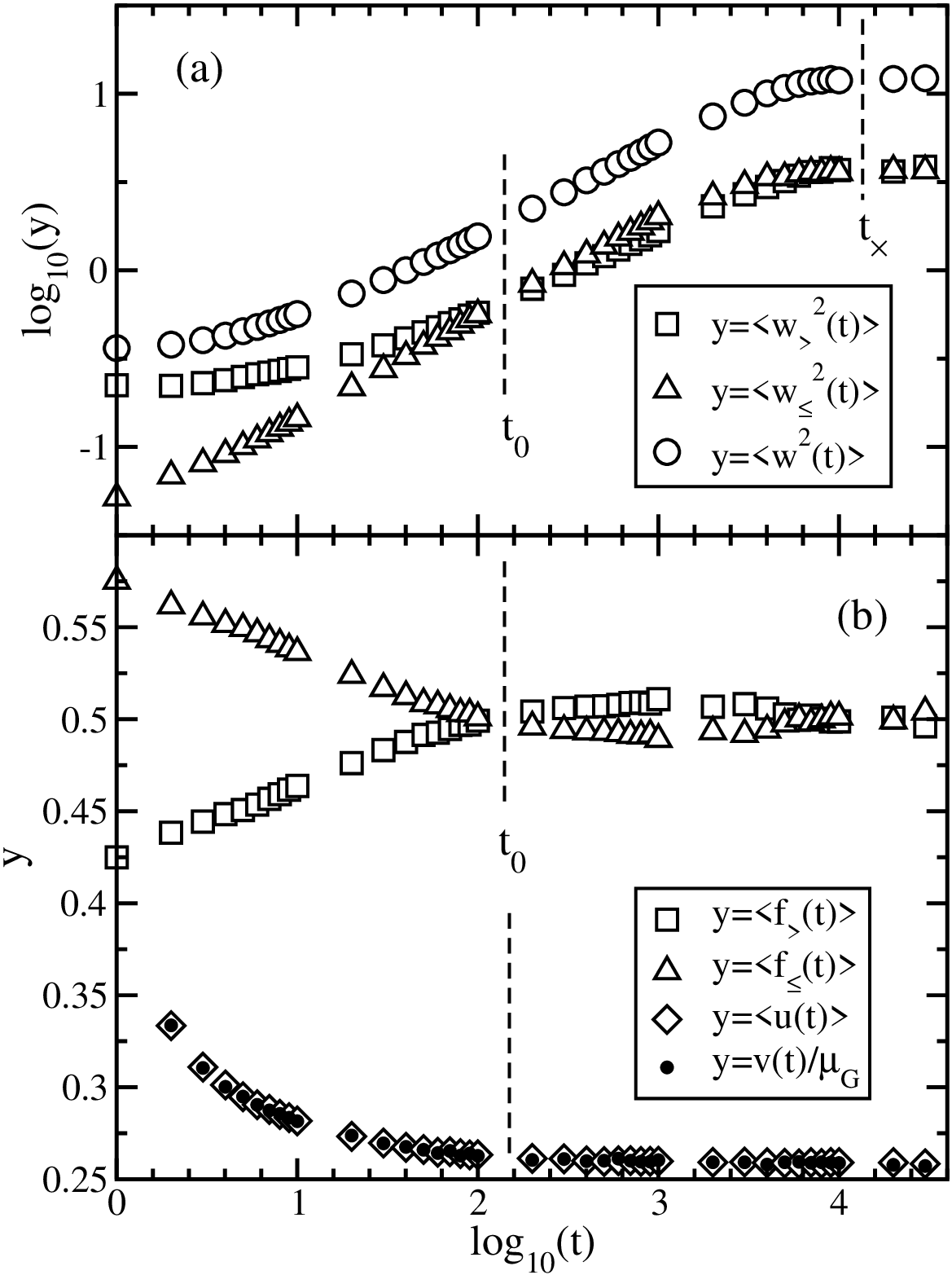}
\caption{\label{pre04f04}
Time evolutions for $N=1$ with Gaussian deposition $(L=1000)$.
(a) The widths, $w^2 = f_{\le}w_{\le}^2 + f_{>}w_{>}^2$.
(b) The interface velocity $v(t)$ and characteristic densities
in analogy with Fig.~\ref{pre04f03}. Here, $\mu _G = \sqrt{2/\pi}$.
}
\end{figure}

\begin{figure}[tp]
\includegraphics[width=6.6cm]{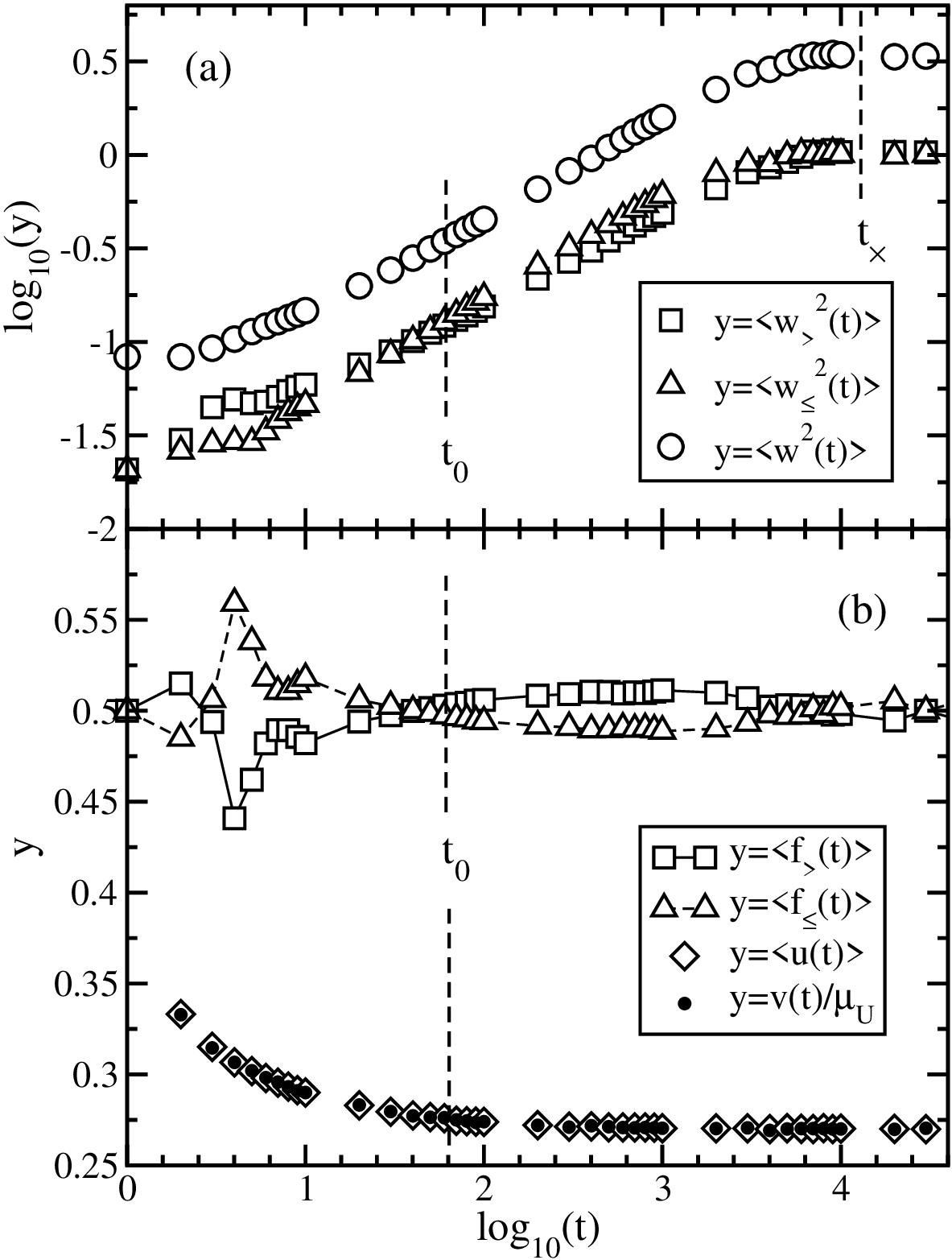}
\caption{\label{pre04f05}
Time evolutions for $N=1$ with uniform deposition $(L=1000)$.
(a) The widths.
(b) The interface velocity $v(t)$ and characteristic densities
in analogy with Fig.~\ref{pre04f03}. Here, $\mu _U = 1/2$.
}
\end{figure}

The initial transition period $t_0$ is an artifact of the flat-substrate initial
condition. To investigate it further, we write out
$w^2(t)$ in its simplex form \cite{KNK03} as the convex linear combination
\begin{equation}
\label{simplex}
w^2(t) = f_{\le}(t) \, w_{\le}^2 (t) + f_{>}(t) \, w_{>}^2 (t) ,
\end{equation}
where $f_{\le}(t) + f_{>}(t)=1$ is the convex sum, i.e., 
$0 \le f_{\le}(t), f_{>}(t) \le 1$. 
The characteristic densities $f_{\le}(t)$ and $f_{>}(t)$
are the fractions of sites that have their heights 
less-then-or-equal-to and larger-than, respectively, 
the mean height $\bar{h}(t)$. The corresponding widths, computed on subsets that
consist of these sites alone, are $w_{\le}^2 (t)$ and $w_{>}^2 (t)$, respectively. 
In individual simulations Eq.(\ref{simplex}) is strictly satisfied and it is valid 
when averaged over many independent simulations. The convex sum is also valid for 
configurational averages of characteristic densities. However, Eq.(\ref{simplex}) 
does not need to hold when characteristic densities and widths are changed to their 
corresponding configurational averages (because, in general, 
$\langle a b \rangle \ne \langle a \rangle \langle b \rangle$). 
Configurational averages of characteristic widths and densities, and the interface 
velocity $v(t)$ are presented in Fig.~\ref{pre04f03}
for Poissonian depositions, in Fig.~\ref{pre04f04} for Gaussian depositions, 
and in Fig.~\ref{pre04f05} for uniform depositions. At $t=0$, the interface
velocity and the utilization have their highest values $\langle u(0) \rangle=1$
because Eq.~(\ref{rule1}) is satisfied at all sites.
This first step at $t=0$ is simply a random deposition step. The mean height
$\bar{h}(0)=\mu$ is the mean of the distribution from which $\eta$ is sampled,
which is $\mu_P =1$, $\mu_G=\sqrt{2/\pi}$ and $\mu_U=1/2$ for Poisson,
Gaussian and uniform depositions, respectively. The fraction 
$f_{>}(0)$ of sites that have their heights larger than $\bar{h}(0)$ is easily computed
from the corresponding distributions as the probability of selecting a site that
has $h(0)$ larger than $\mu$. This gives for Poisson deposition
$f_{>}(0)=\int_{\mu_P}^{\infty}dx \exp(-x) =1/e \approx 0.367$;
for Gaussian deposition
$f_{>}(0)=\sqrt{2/\pi}\int_{\mu_G}^{\infty}dx \exp(-x^2/2) =
1- \mathrm{erf}(1/\sqrt{\pi}) \approx 0.428$; and, for uniform deposition
$f_{>}(0)=\int_{\mu_U}^{1}dx =1/2$. These fractions and their complements
$f_{\le}(0)=1-f_{>}(0)$ are clearly observed in Figs.~\ref{pre04f03}-\ref{pre04f05}.
Correlations between lattice sites start to build up at $t=1$.
Since initially the density $f_{\le}(t)$ is larger than $f_{>}(t)$,
depositions take place more often at sites with $h \le \bar{h}$ then at sites with
$h > \bar{h}$. This causes $f_{\le}(t)$ to fall and $f_{>}(t)$ to rise
(Figs.~\ref{pre04f03}{\it b}-\ref{pre04f05}{\it b}) and a faster growth of 
$w_{\le}^2(t)$ than $w_{>}^2(t)$ (Figs.~\ref{pre04f03}{\it a}-\ref{pre04f05}{\it a}).
On the average, as the density $\langle f_{>}(t) \rangle$ rises the density of local
minima $\langle u(t) \rangle$ decreases. This initial evolution from the RD 
surface at $t=0$ to a surface with correlations at $t_0$ ends when
$\langle f_{>}(t) \rangle \approx \langle f_{\le}(t) \rangle \approx 1/2$. As
Figs.~\ref{pre04f03}-\ref{pre04f05} illustrate, at $t_0$ the simulations attain 
what we label as a 
steady state, one that is characterized by a constant utilization. We show in
Sec.~\ref{velocity} that $v(t)$ is related to $\langle u(t) \rangle$ 
by a simple linear relation, hence, the steady state can be
alternatively defined by a constant velocity.

\begin{figure}[bp]
\includegraphics[width=8.4cm]{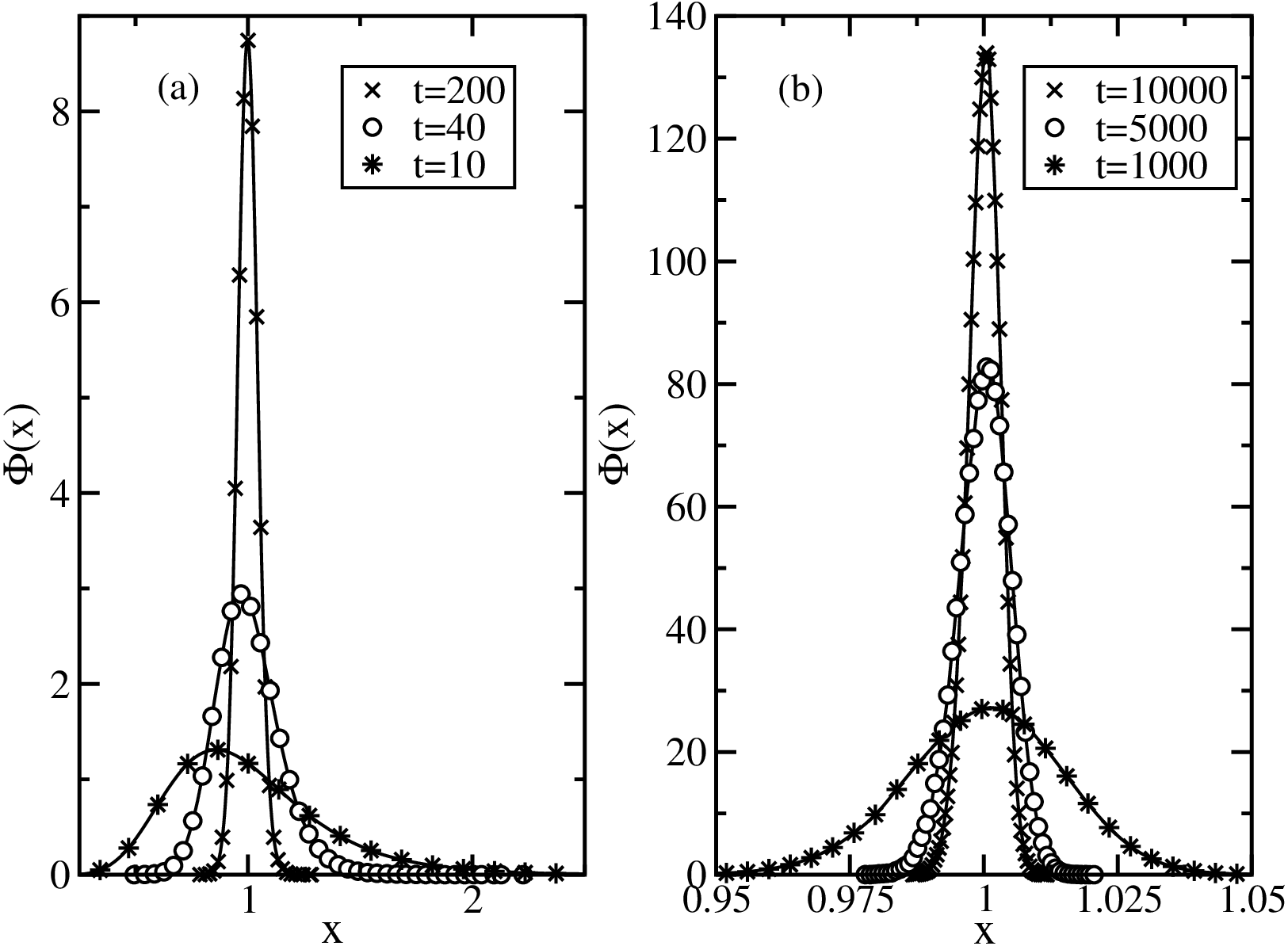}
\caption{\label{pre04f06}
Poisson deposition at local surface minima for $L=1000$: distribution function
$\Phi(x)$ of the interface local heights $x=h(t)/\bar{h}(t)$. (a) Early times
$t \le t_0$. (b) The steady state for $t > t_0$. Cubic-spline curves
through the simulation data (symbols) are guides for the eyes. Here, $t_0 \propto 100$.
}
\end{figure}

The correlated growth phase when the scaling is observed, 
when $t_0 < t \ll t_{\times}$, is characterized by 
a slight but noticeable excess of $\langle f_> (t) \rangle$ over $\langle f_{\le} (t) \rangle$. 
At saturation, for
$t \gg t_{\times}$, $\langle f_{>} \rangle \approx \langle f_{\le}\rangle$.
The densities $\langle f_{\le}(t) \rangle$ and $\langle f_{>}(t) \rangle$,
and the widths $\langle w_{\le}^2(t) \rangle$ and $\langle w_{>}^2(t) \rangle$,
provide the information about the height distribution $\Phi (h/\bar{h})$ of the
interface local heights about the mean height $\bar{h}(t)$. It is transparent
from Figs.~\ref{pre04f03}-\ref{pre04f05} that for early times, $t<t_0$, $\Phi (h/\bar{h})$ is
characterized by a positive skewness and evolves to approximately a symmetric
distribution at $t_0$. This distribution function for Poisson depositions at
local minima is presented in Fig.~\ref{pre04f06}. The computation of $\Phi (x)$
is outlined in the Appendix.

The skewness of the height distribution in the stationary state of the KPZ growth 
has been analyzed before by den Nijs and co-workers \cite{Nijs00,NN97,CN99} for 
Kim-Kosterlitz models with integer step-height differences. They report that KPZ 
scaling is realized at times larger than a characteristic time scale that is related 
to slope densities. In (1+1) dimension the KPZ dynamics is characterized by zero 
skewness because the height distribution of the stationary state is Gaussian \cite{NN97}. 
In our models, the growth can be characterized alternatively either by the density of 
local minima (i.e., the utilization for $N=1$) that is the same as the density of local 
maxima \cite{KNR03} or by the density of local slopes. Since all these densities sum up 
to one, the constant utilization in our model is equivalent to a constant density of local 
slopes. Explicitly, we define the steady evolution state (or the steady state simulations) 
as the evolution that has the following characteristics: 1. the density of update sites 
is constant; and, 2. the skewness of the height distribution is approximately zero. 
Starting from the flat substrate, the steady growth state is achieved after the initial 
relaxation time $t_0$. In the steady state the KPZ scaling is clearly observed.

The initial time interval from $t=1$ to $t=t_0$ can be interpreted as the time
scale over which the system retains the memory of the flat-interface initial condition.
This time is a nonuniversal parameter that depends on the variance of the distribution
from which the random height increment $\eta$ is sampled. The existence of this time
scale accounts for the absence of universal scaling for small system sizes,
even if the rule that simulates the growth represents a generic KPZ process. For
KPZ dynamics the characteristic time scale on which the correlations
are being built is of the order of the system size $t_{\times} \sim L^{3/2}$.
If this time scale is smaller than the memory scale, $L^{3/2} < t_0$, the interface
saturates before the simulations reach the steady state, and for
such $L$ the KPZ scaling is not observed. The universal KPZ scaling is clearly observed
when the system size is large enough to loose the memory of the initial condition on
time scales smaller than $L^{3/2}$, i.e., when $t_0 \ll L^z$.

\section{SCALING ANALYSIS \label{scaling}}

In this section we analyze the interfaces generated by deposition rule $3$ that
represents two simultaneously acting growth mechanisms: one is
RD and the other is deposition rule $2$, both with Poisson-random height increments.
First we show that rule $2$ generates KPZ correlations. Then we obtain the
universal scaling function for interfaces produced by rule $3$.

\begin{figure}[tp]
\includegraphics[width=6.6cm]{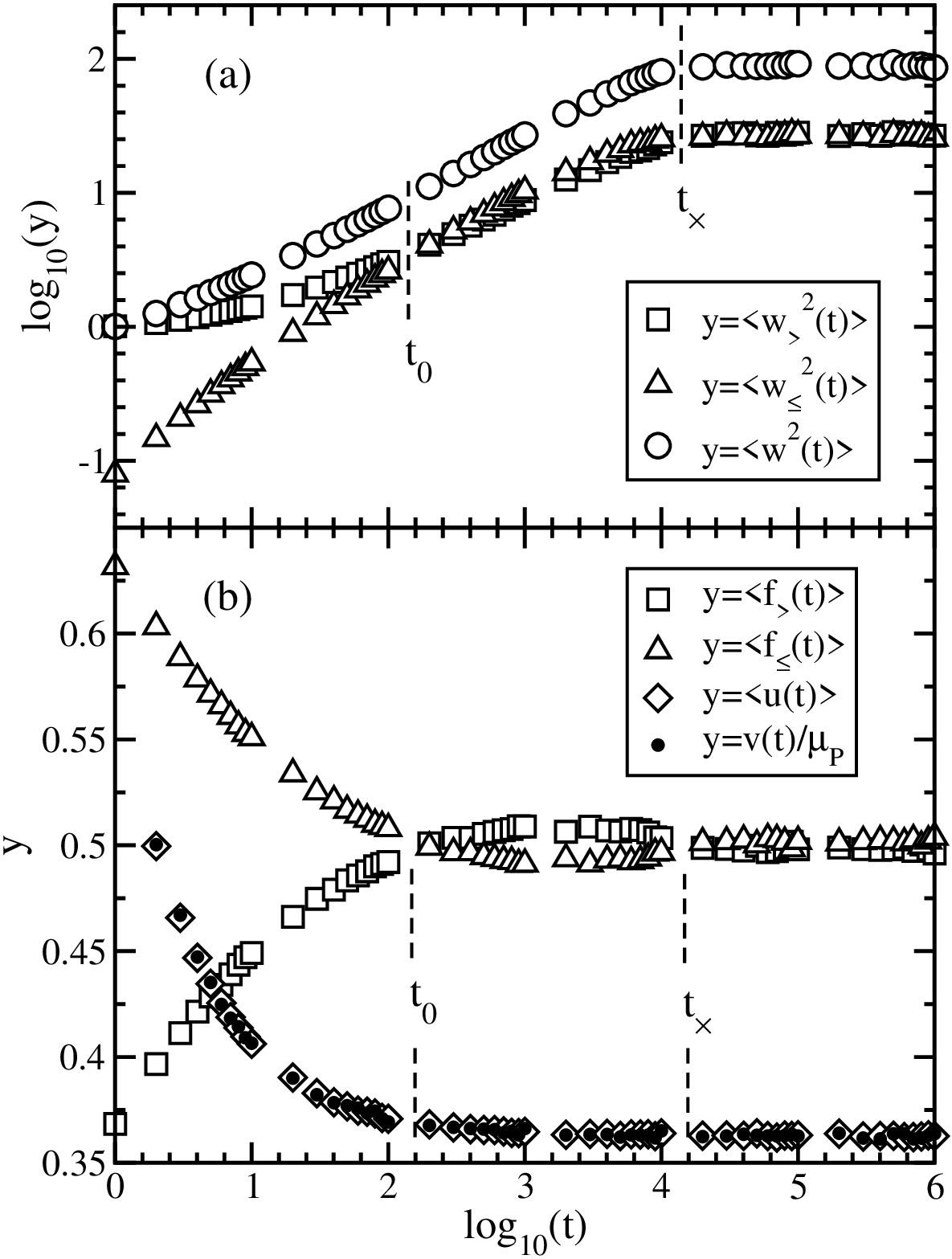}
\caption{\label{pre04f07}
Time evolutions for Poisson deposition when $N=2$ and $L=1000$.
(a) The widths. (b) Characteristic densities and the interface velocity,
in analogy with  Fig.~\ref{pre04f03}.
}
\end{figure}
\begin{figure}[tp!]
\includegraphics[width=6.4cm]{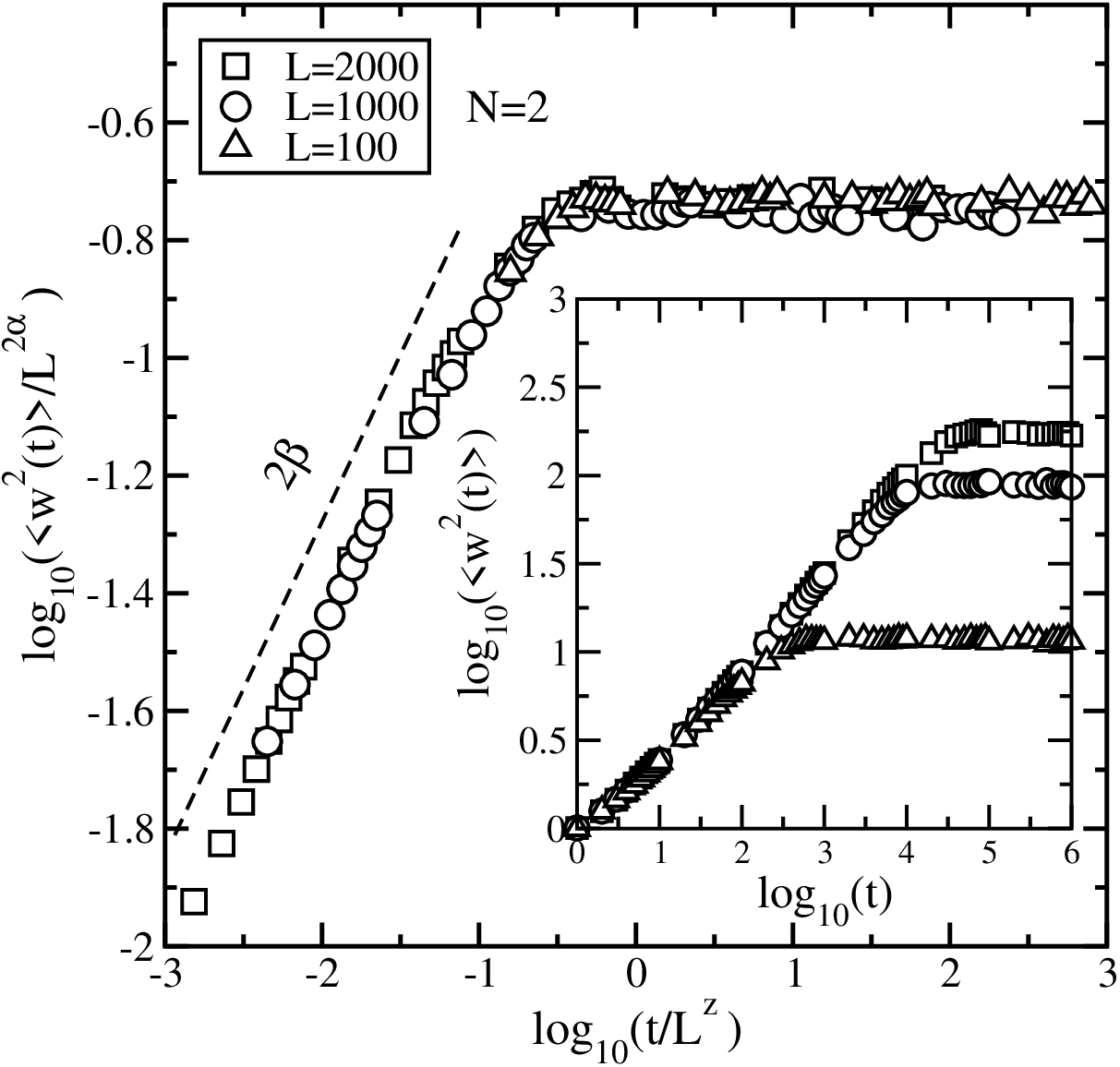}
\caption{\label{pre04f08}
Scaling for $t>t_0$ when $N=2$. The initial relaxation period for
$t<t_0$ is clearly noticeable in the insert that shows the widths before scaling.
}
\end{figure}

Although rule $2$ allows the $k$th site to accept a deposition even if it is
not a local minimum, this rule has all the essential characteristics of rule $1$, examined
in Sec.~\ref{flat}. At each time step a site must compare its local height with a local
height of at least one of its immediate neighbors. As in rule $1$, deposition may not
happen at a local maximum. But, since now it may happen either at a local minimum or at
a local slope, the utilization of rule $2$ is larger than the one of rule $1$
(compare Fig.~\ref{pre04f03}{\it b} and Fig.~\ref{pre04f07}{\it b}) so the interface velocity is
larger. Other than that there is no difference between these two deposition mechanisms, 
and the analysis presented in Sec.~\ref{flat} for interfaces produced by rule $1$ can
be restated for the interfaces that grow by rule $2$. In particular, both growths evolve
on the same time scales (Fig.~\ref{pre04f07}), with the initial memory scale
$t_0 \propto 100$. Figure~\ref{pre04f08} shows the scaling function for the interface
widths for $t>t_0$, obtained with $2\alpha = 0.9$ and $z=2-\alpha$, characteristic
for KPZ scaling. Thus, these interfaces belong to the KPZ universality class and
the scaling function is given by Eqs.~(\ref{Family1}-\ref{Family2}). A small departure  
of $2\alpha$ from one, also present for $N=1$, is discussed in Sec.~\ref{roughness}.

\begin{figure}[tp]
\includegraphics[width=6.4cm]{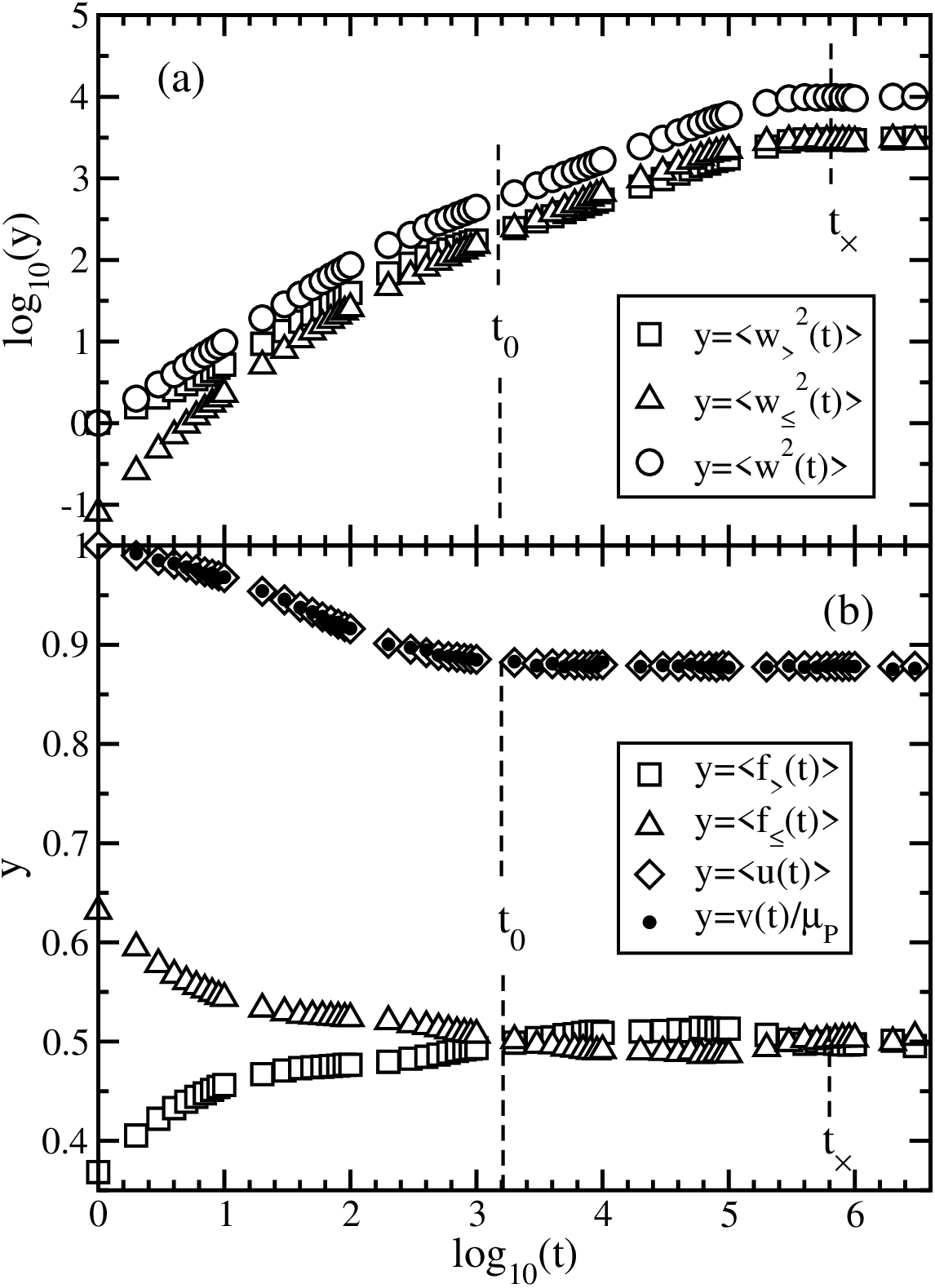}
\caption{\label{pre04f09}
Time evolutions for Poisson deposition when $N=100$ and $L=1000$.  
(a) The widths. (b) Characteristic densities and the interface velocity, 
in analogy with  Fig.~\ref{pre04f03} and Fig.~\ref{pre04f07}.
}
\end{figure}

\begin{figure}[tp!]
\includegraphics[width=6.0cm]{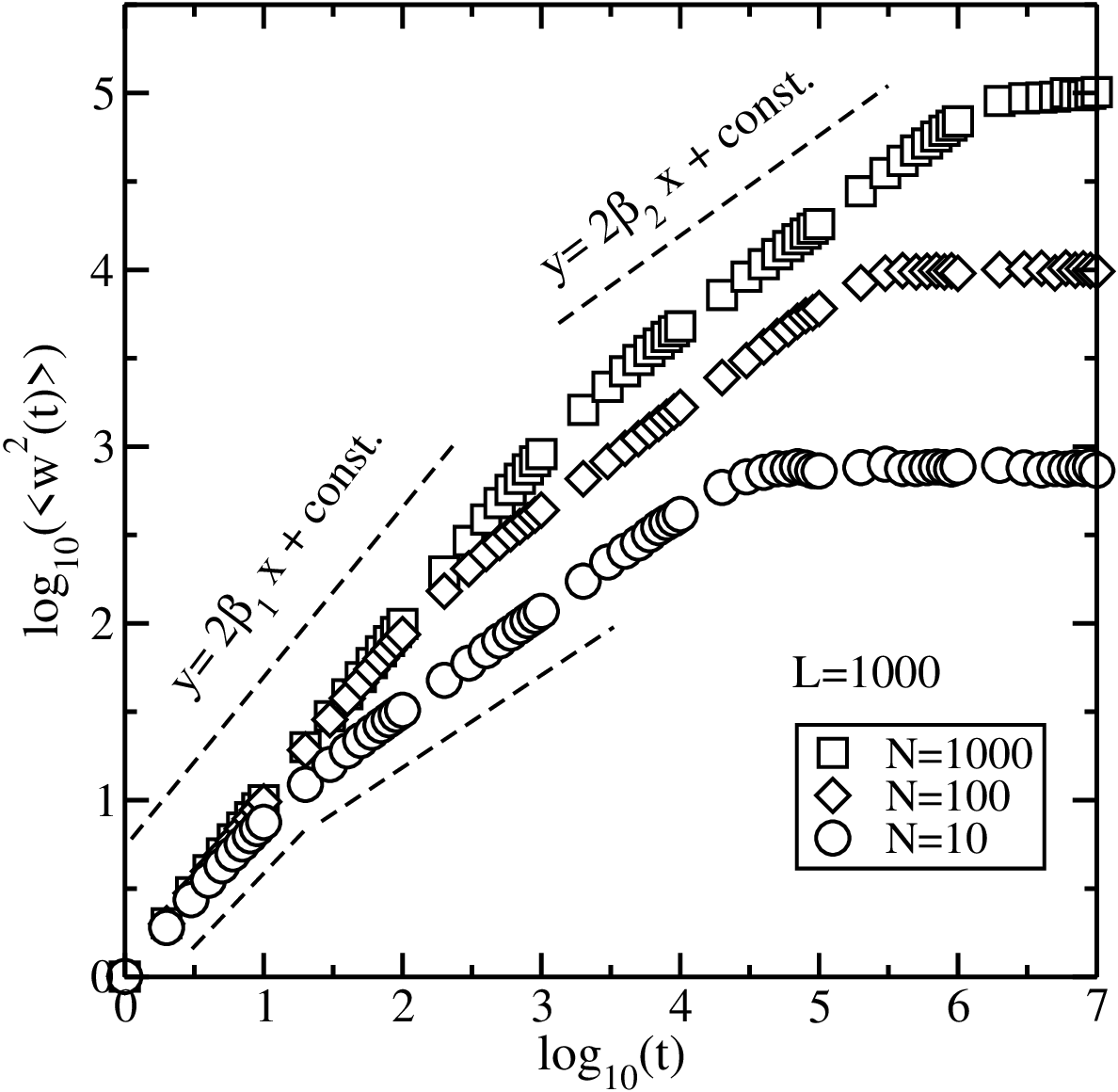}
\caption{\label{pre04f10}
Two growth phases in the time evolution of the width, simulated with
rule $3$ for $N>3$ and $L=1000$.
}
\end{figure}

Deposition rule $3$ produces a larger utilization than rule $2$ because now,
depending on $N$, deposition at site $k$ may sometimes be accepted
regardless of the relation between its local height and local heights of its neighbors.
Now at each $t$, any site may increase its height, including a local maximum.
Probability $p(N)$ that a site has to compare its local height with a neighbor is
the probability of applying the rule $2$, which is the only mechanism that creates
correlations. Alternatively, deposition at the $k$ site may
happen as RD. A combination of these two deposition mechanisms produces
a similar time evolution of characteristic densities and the widths as it is observed
when rule $2$ is acting alone, except that now the transition to steady-state simulations 
and the cross-over to saturation take place on larger time scales
(compare Fig.~\ref{pre04f07} and Fig.~\ref{pre04f09}). In particular, the initial
lack of scaling extends to $t_0 (N) \propto t_0 N/2$, where $t_0$ marks the end 
of the initial relaxation period in the worst-case scenario simulations. 
This initial relaxation time $t_0(N)$, 
when the system ``remembers" the flat-interface initial condition, manifests itself
in the evolution of interface widths as an early growth phase (Fig.~\ref{pre04f10})
that follows the RD power law with $\beta_1 =1/2$. The later growth phase,
$t_0(N) \ll t \ll t_{\times}$, follows the power law with
$\beta_2 =1/3 - \varepsilon$, where $\varepsilon$ is a small positive number.
The evolution of the interface width can be summarized as
\begin{equation}
\label{evolution}
\langle w^2 (t) \rangle \sim \left\{ \begin{array} {r@{\quad , \quad}l}
t^{2 \beta_1} & t < t_0 (N) \\
t^{2 \beta_2} & t_0 (N) \ll t \ll t_{\times} (N) \\
g(N) L^{2\alpha} & t \gg t_{\times}(N) ,  \end{array} \right.
\end{equation}
where $g(N)$ is a monotonic function of $N$. After performing scaling in $L$ of the
saturated width, it appears that $g(N)$ is linear. The first growth phase
is the initial relaxation when $\langle w^2 (t) \rangle$ does not scale,
therefore the following analysis is valid only for $t > t_0 (N)$.

\begin{figure}[tp!]
\includegraphics[width=8.5cm]{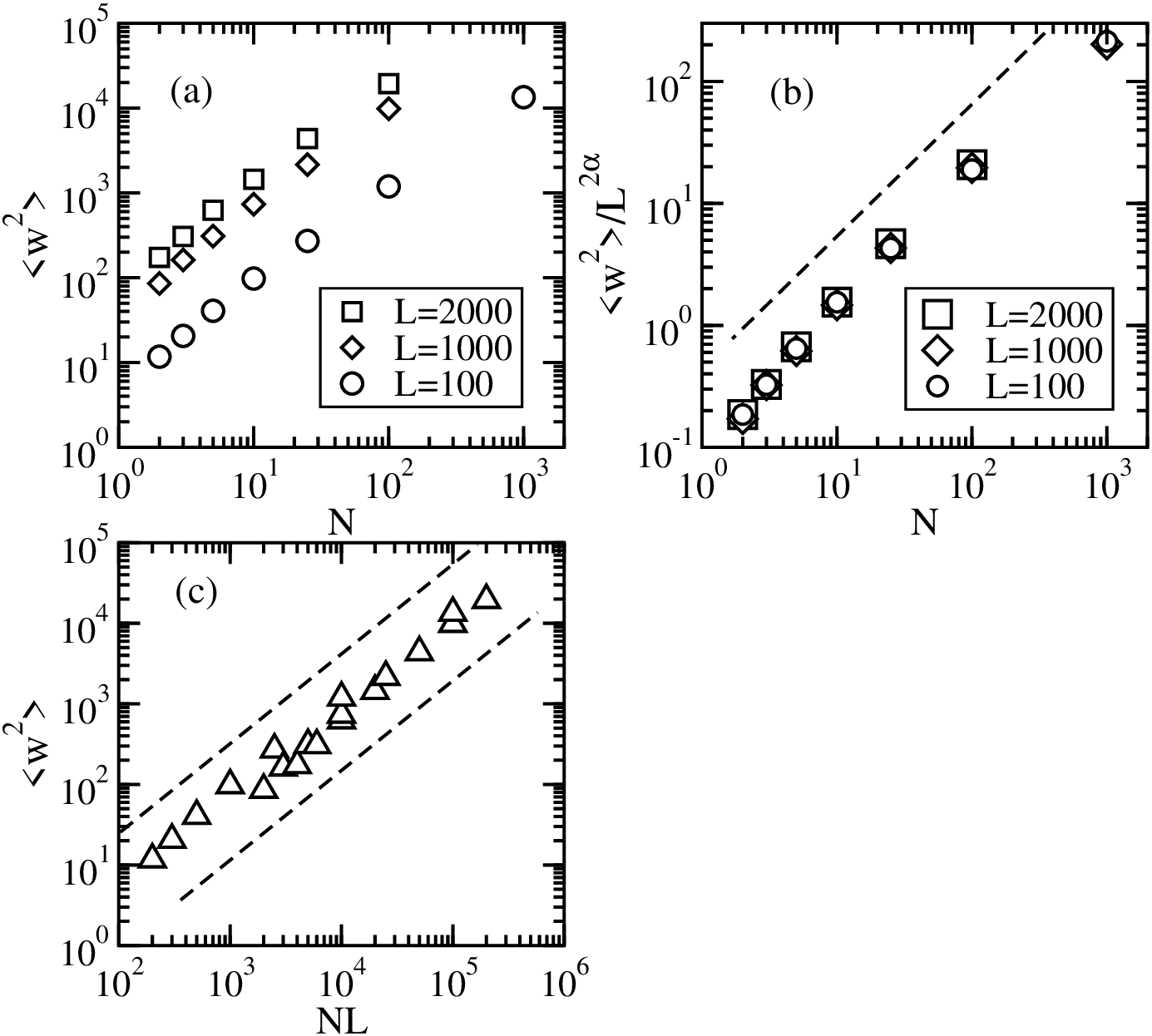}
\caption{\label{pre04f11}
The saturated interface width as a two-parameter family of curves obtained in 
simulations with rule $3$: (a) as a function of $N$ volume elements 
per lattice site for several lattice sizes $L$; 
(b) as a function of $N$, after scaling in $L$ of the data in figure 
(a) (the dashed line of slope one is plotted as a reference); 
(c) data from figure (a) plotted vs $(NL)$ to see global trends.  
The data align along a straight line of the mean slope one (the dashed-line envelope).
}
\end{figure}

From the point of view of scaling, $\langle w^2 (t) \rangle$ is a family of curves
parametrized by $L$ and $N$. Figure~\ref{pre04f11}{\it a} presents the saturated width
$\langle w^2 \rangle$ plotted against $N$ for selected values of $L$. These curves
can be scaled in $L$ so as they collapse onto one curve. Figure~\ref{pre04f11}{\it b} shows
the scaled width $\langle w^2 \rangle /L^{2\alpha}$, where $2\alpha=0.9$.
Since $\log_{10}({\langle w^2 \rangle /L^{2\alpha}}) \sim \log_{10}{N}+ \mathrm{const.}$
(dashed line in Fig.~\ref{pre04f11}{\it b}), values at saturation may be
further scaled in $N$, which gives the  collapse to one point
$\langle w^2 \rangle /(NL^{2\alpha}) \approx \mathrm{const}$. The order
of scaling can be reversed, i.e., scaling in $N$ can be followed by scaling in $L$,
leading to the same result. Globally, the saturated width plotted vs $(NL)$ follows
a line of slope one (Fig.~\ref{pre04f11}{\it c}).

\begin{figure}[tp!]
\includegraphics[width=8.5cm]{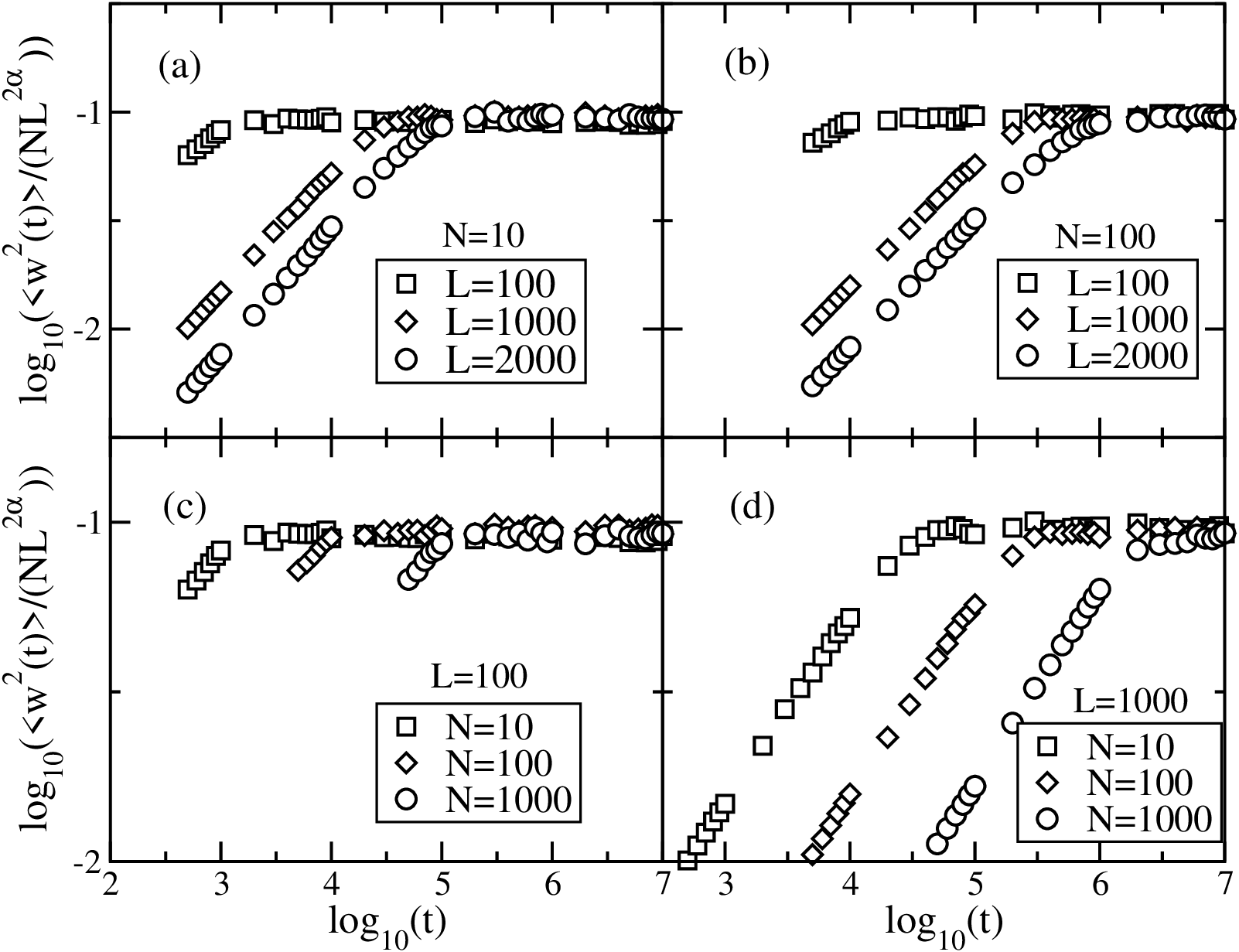}
\caption{\label{pre04f12}
The steady-state evolution for $t>t_0$ of the scaled widths obtained with 
rule $3$ as a two-parameter family of curves for: (a) $N=10$; (b) $N=100$;
(c) $L=100$; and, (d) $L=1000$.
}
\end{figure}

\begin{figure}[tp]
\includegraphics[width=7.0cm]{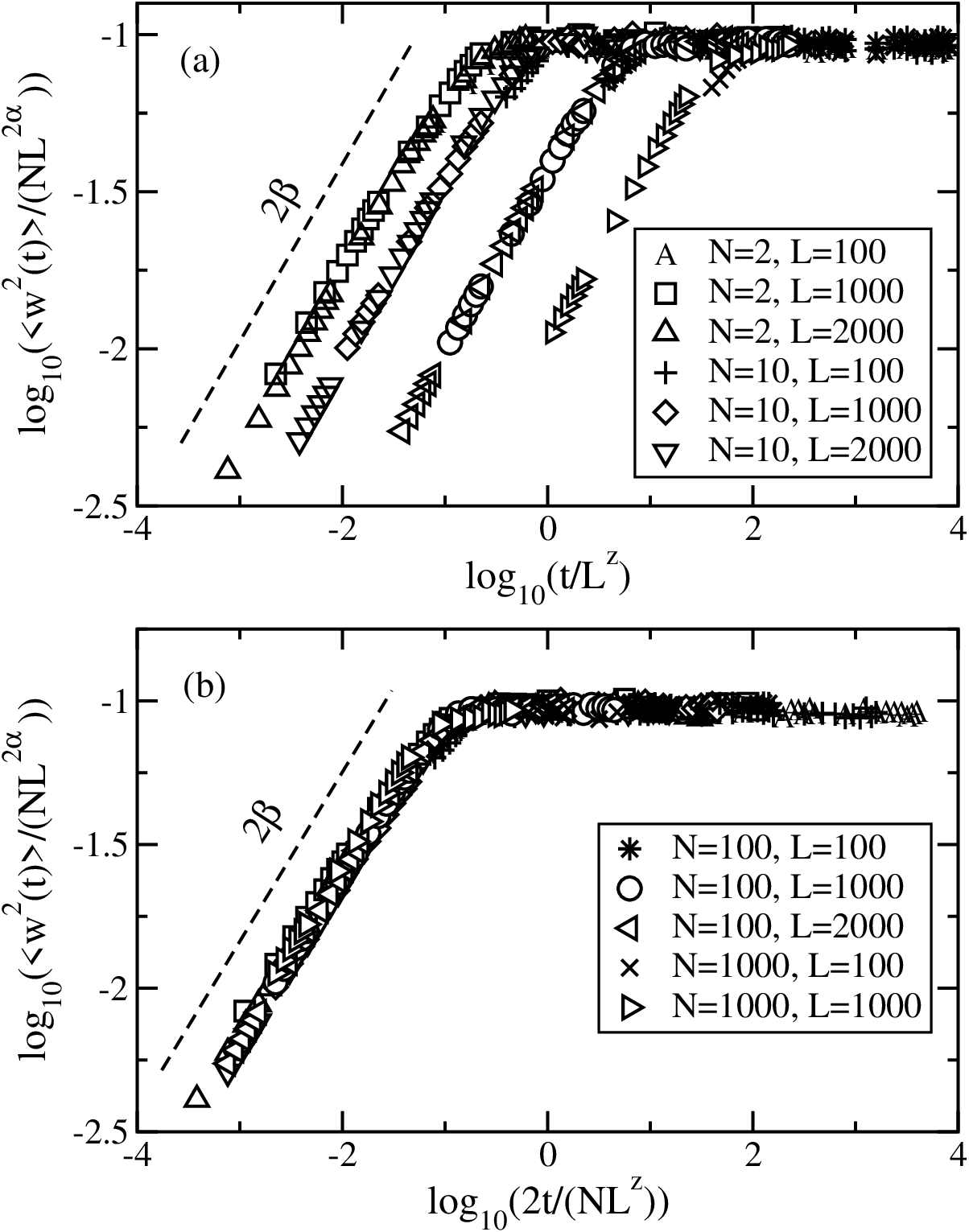}
\caption{\label{pre04f13}
Scaling for the curves of Fig.~\ref{pre04f12}:
(a) in $t$ with respect to $L$, with the KPZ dynamic exponent
$z=2-\alpha$; (b) in $t$ with respect to $N$ of the data in figure (a).
Data labels displayed in the legend are common for both figures (a) and (b).
}
\end{figure}

Time evolution of the scaled width $\langle w^2 (t) \rangle /(NL^{2\alpha})$ for 
$t>t_0(N)$ is displayed in Fig.~\ref{pre04f12}. Here, to obtain a perfect align 
at saturation with the curves for $N=2$, which facilitates further scaling in $t$, 
we introduced a small correction such that $NL^{2\alpha}$ is multiplied by 
$(1 \pm \varepsilon)$, where $\varepsilon$ is a small fraction (explicitly, 
$\varepsilon$ is the relative spread of data about the fit at saturation; 
the scaling is clearly seen with $\varepsilon =0$). Since the only 
mechanism that induces correlations in this model is deposition rule $2$, 
which  produces surfaces from the KPZ universality class, it is expected that 
for each $N$ the length scales should couple with time scales via a dynamic exponent 
that satisfies the KPZ identity $z=2-\alpha$. Indeed, as the partial scaling in $t$ 
with respect to $L$ shows (Fig.~\ref{pre04f13}{\it a}), this choice gives 
a good data collapse for each $N$ 
and this is not changed noticeably when $z=1.5$ is chosen.  
In Fig.~\ref{pre04f13}{\it a},  
reading from the left, the groups represent $N=2$, $N=10$, $N=100$, and $N=1000$. 
The last step is the scaling in $t$ with respect to $N$ of the results displayed in 
Fig.~\ref{pre04f13}{\it a}. Here the inspection of the dilatation with $N$ of the initial 
$t_0$ proves useful since it leads to the observation that for any $L$ the initial $t_0(N)$ 
can be expressed approximately as $(N/2)t_0(N=1 \; \mathrm{or}\; 2)$. The transformation 
$t \to t/(N/2)$ shifts (to the left) all the curves for $N>2$ into one position. 
The family of curves for $N=2$ is shifted into this position when $t \to t/2$. 
The final result is the scaling function shown in  Fig.~\ref{pre04f13}{\it b}. 
The mean slope of the growth part is $2\beta \approx 0.58 \pm 0.02$, consistent with 
the slope obtained for $2\alpha =0.9$ from the KPZ relation 
$2\beta= (2\alpha)/(2-\alpha) \approx 0.58$.

The final result, valid for $t>t_0(N)$, can be summarized as
\begin{equation}
\label{final}
\langle w^2(t) \rangle = \frac{N}{2} L^{2\alpha} f \left(\frac{2}{N}\frac{t}{L^z}\right),
\end{equation}
where $f(y)$ satisfies Eq.~(\ref{Family2}), and $z=2-\alpha$ with $\alpha \approx 1/2$. 
Accordingly, the interfaces generated by the deposition/update rule $3$ belong to 
the KPZ universality class. In the scaling regime, the evolution can be written 
out explicitly as
\begin{equation}
\label{explicit}
\langle w^2 (t) \rangle \sim \left\{ \begin{array} {l@{\quad , \quad}l}
(\frac{N}{2})^{1-2\beta} \, t^{2 \beta} & t_0 (N) \ll t \ll t_{\times} (N) \\
\frac{N}{2} L^{2\alpha} & t \gg t_{\times}(N) ,  \end{array} \right. 
\end{equation}
where $t_{\times}(N) \sim (N/2) L^z$.

\section{INTERFACE VELOCITY \label{velocity}}

For the deposition/update models considered in this work it is possible to find 
the exact relation between the utilization and the interface velocity. 
The velocity $v(t)$ is defined as the configurational average of 
$\bar{v}(t) = d \bar{h}(t)/dt = (1/L) \sum_{k=1}^L d h_k(t)/dt$. 
Translating the update recipe to a continuum version 
with a continuum time-step increment $\delta t$, the update 
operation at site $k$ can be summarized as
\begin{equation}
\label{update}
h_k (t+\delta t) = \left\{ \begin{array} {l@{\quad , \quad}l}
h_k(t) + \delta t \, \eta _k (t) & \mathrm{on} \;\, \mathrm{update} \\
h_k(t) & \mathrm{otherwise}  .  \end{array} \right. 
\end{equation}
Substituting the above to the definition of $d h_k(t)/dt$ as the limit when 
$\delta t \to 0$, gives at $t$
\begin{equation}
\label{derivative}
\frac{d h_k (t)}{dt} = \left\{ \begin{array} {l@{\quad , \quad}l}
\eta _k (t) & k\;\,\mathrm{is} \;\,\mathrm{the} \;\,\mathrm{update} \;\,\mathrm{site} \\
0 & \mathrm{otherwise} .  \end{array} \right. 
\end{equation}
In the set of $L$ sites the number of update sites is $M(t)=L u(t)$. 
Since only at these sites $d h_k(t)/dt$ is not zero, the mean is
\begin{equation}
\label{mean}
\bar{v}(t) =\frac{1}{L} \sum_{k=1}^{M(t)} \eta_k(t) 
= u(t) \frac{1}{M(t)} \sum_{k=1}^{M(t)} \eta_k(t) .
\end{equation}
Let $\mu _P$ be the mean of the distribution $P(\eta)$ from which $\eta _k$ is sampled. 
In the limit of $M \to \infty$, 
$(1/M) \sum_{k=1}^{M} \eta _k \to \int_{\Omega} d \eta \, \eta \, P(\eta) = \mu_P$. 
Thus, for sufficiently large $M$, the second factor in Eq.~(\ref{mean}) 
is $\mu _P = \mathrm{const}$. 
Taking the configurational average of Eq.~(\ref{mean}) gives 
\begin{equation}
\label{relation}
v(t) = \langle u(t) \rangle \, \mu _P .
\end{equation}
The above derivation can be repeated for the discrete case, taking $\delta t =1$. 
Equation~(\ref{relation}) is strictly satisfied by the simulation data 
(Figs.~\ref{pre04f03}-\ref{pre04f05}, \ref{pre04f07} and \ref{pre04f09}). 
In simulations $\bar{v}(t)$ is computed numerically with 
$\delta t =1$, $\bar{v}(t) = \bar{h}(t) - \bar{h}(t-1)$, and $v(t)$ is obtained 
by averaging $\bar{v}(t)$ over many independent simulations.

\section{DISCUSSION \label{discussion}}

In our steady-state simulations for $N>1$, the mean density 
of sites $p(N)=\sqrt{2/N}$ can be interpreted as the probability 
that at $t$ a randomly selected site followed rule $2$.
The complementary density $\bar{p}=1-p$ 
is the probability that a randomly selected site followed RD. A similar interpretation 
can be given to the mean density of update sites 
$\langle u(t) \rangle$ as the probability that at $t$ a randomly selected lattice 
site increased its height; however, $\langle u(t) \rangle$ does not define a probability 
distribution \cite{KNR03,KN04}. Using a recently introduced discrete-event analytic 
technique for surface growth problems \cite{KNR03,KN04}, it is possible to derive a 
mean-field-like expression for the utilization $\langle u \rangle$ in the steady-state 
simulations. For $N \ge 2$ and $L \ge 3$ this expression \cite{KN04} is
\begin{equation}
\label{steadyu}
\langle u \rangle = \left( 1- \frac{p(N)}{2} \right) 
\left( 1- \frac{p(N)}{4} \frac{L-1}{L}  \right).
\end{equation}
Because $\langle u(t) \rangle$ and $v(t)$ are related by a constant multiplicative 
factor $\mu _P$, by Eq.~(\ref{relation}), in the scaling regime for $t>t_0(N)$, 
Eq.~(\ref{steadyu}) also gives the interface velocity. However, for early times 
$\langle u(t) \rangle$ is unknown.

\subsection{Kinetic roughening \label{roughening}}

The scaling expressed by Eq.~(\ref{final}) can be written in a more general 
form by incorporating the probability $p=\sqrt{2/N}$. This gives
\begin{equation}
\label{final2}
\langle w^2(t) \rangle = \left( \frac{L^{\alpha}}{p} \right)^2 f\left(\frac{p^2t}{L^z}\right) .
\end{equation}
The analysis of Sec.~\ref{scaling} can be repeated for $\langle w(t) \rangle$, 
which gives equivalently
\begin{equation}
\label{final3}
\langle w(t) \rangle =  \frac{L^\alpha}{p} \, f \left(\frac{p^2t}{L^z}\right) .
\end{equation}
The transition time to the scaling regime is $t_0 (p) = t_0 /p^2$, where $t_0$ marks 
the end of the initial non-scaling period in the case of simulations with the 
deposition/update rule $2$ acting  alone. The cross-over time to saturation 
$t_\times$ can be read directly from Eqs.~(\ref{final2}-\ref{final3}), 
$p^2 t_\times (p) /L^z \approx 1$. This gives $t_\times (p) \approx t_\times/p^2$, 
where $t_\times$ is the saturation time when rule $2$ acts alone.

Our results for the scaling function, Eqs.~(\ref{final}-\ref{explicit}) and 
Eqs.~(\ref{final2}-\ref{final3}), are generally in accord with the work of Horowitz and 
Albano \cite{HA01} and Horowitz {\it et al.} \cite{HMA01}, who analyzed (1+1) 
dimensional two-component solid-on-solid models mixed with RD. In Ref.~\cite{HA01}, 
the growth is simulated by ballistic deposition (of the KPZ universality class) 
that takes place with probability $p$ and by RD that happens with probability $(1-p)$. 
In Ref.~\cite{HMA01} the deposition model mixes RD (taking place with probability $(1-p)$) 
and random deposition with surface relaxation (of the EW universality class) that takes 
place with probability $p$. A common characteristic of these two models and our model 
is the presence of two growth phases in time evolution of the interface width 
(Fig.~\ref{pre04f10} and  Eq.~(\ref{evolution})), where the early phase follows RD 
growth and this is the phase that defies the universal scaling 
(figures 4 in Refs.~\cite{HA01,HMA01}). Although this initial absence of scaling is 
not studied in Refs.~\cite{HA01,HMA01}, based on our analysis of Sec.~\ref{flat}, 
we infer that this early RD growth seen in Refs.~\cite{HA01,HMA01} 
must be a long-time effect of some particular initial condition 
(not stated in \cite{HA01,HMA01}) adopted in these simulations. As we showed in 
Sec.~\ref{flat}, the length of this initial memory scale is a 
nonuniversal parameter, so comparing cross-over times to the scaling regime does not 
contain useful information. But, the existence of this initial memory scale and the 
existence of scaling for times larger than the initial relaxation time are both universal. 
The scaling law reported in Refs.~\cite{HA01,HMA01} is
\begin{equation}
\label{theirs}
W(t, L, p) \propto \frac{L^{\tilde{\alpha}}}{p^{\delta}} \, 
F\left(\frac{p^yt}{L^{\tilde{z}}}\right) , 
\end{equation}
where the exponents $\tilde{\alpha}$ and $\tilde{z}$ are characteristic for the 
universality class of the model (i.e., KPZ in \cite{HA01} and EW in \cite{HMA01}). 
For the mixture of RD and random deposition with surface relaxation 
(the EW type model) the numerics give $\delta \approx 1$ and 
$y \approx 2$ \cite{HMA01}, in agreement with our findings, Eq.~(\ref{final3}). 
However, for the mixture of RD and ballistic deposition (the KPZ type model) 
it is conjectured in \cite{HA01} that $\delta \approx 1/2$ and $y \approx 1$. 
In this latter scaling, although the relation between $y$ and $\delta$, $y=2 \delta$, 
agrees with our findings, these powers are by the factor of $2$ smaller than 
our values. As we analyze later in this section, this difference implies that according 
to Ref.~  \cite{HA01} the admixture of RD should affect time scales as $pt$, 
while our findings clearly indicate the $p^2 t$ behavior.

The sensitivity of the surface evolution to the initial condition 
has been recently pointed out by Kortla {\it et al.} \cite{KPS98} in relation with 
phase ordering in two-component solid-on-solid models. In our modeling, to incorporate 
fully the dynamics of the growth, the KPZ equation should contain the mean interface 
velocity $v(t)$
\begin{equation}
\label{kpz2}
h_ t = v(t) + \nu \, h_{xx} + \frac{\lambda}{2} \, h_x^2 + \zeta  \,  .
\end{equation}
Equation~(\ref{kpz}) can be valid only for the steady-state simulations, when 
$v(t) = \mathrm{const}$. Therefore, it does not describe the initial dynamics 
at $t<t_0$ of our deposition models, while Eq.~(\ref{kpz2}) does. 
Results obtained in Sec.~\ref{scaling}, 
summarized by Eqs.~(\ref{final2}-\ref{final3}), indicate that the admixture of the 
RD process (present with probability $(1-p)$ in rule $3$) elongates 
time scales in Eq.~(\ref{kpz2}). We now scrutinize this effect from the point of view 
of the affine transformations involved.

Consider the following transformations that are expressed by Eqs.~(\ref{final2}-\ref{final3})
\begin{equation}
\label{trans1}
x \to x'=\frac{x}{L} \, ,\; \; h \to h'= \frac{g(p)}{L^\alpha} h \, , \; \; 
t \to t' = \frac{p^2}{L^z}t \, ,
\end{equation}
where $g(p)$ is to be determined (note, $g(p)$ is not independent because $x$, $h$ 
and $t$ are transformed simultaneously). Denoting $v= - h_x$, in (1+1) dimensions 
the convective  derivative in Burger's flow is $D_t v = v_t + \lambda\,v\,v_x$ \cite{BS95}. 
It is straightforward to show that $D_t v$ is invariant (up to a multiplicative factor) 
under transformations (\ref{trans1}) if $\alpha +z=2$, $g(p)=p$ and if 
$\lambda \to \lambda'$, $\lambda' p^3= \lambda$. Then $w^2(t)$ transforms 
to $w'^2(t')$, $w'^2(t') = w^2(t) / (L^{2\alpha}/p^2) \sim \Phi (t')$ or, equivalently, 
$w(t) \to w(t)/(L^{\alpha}/p) ~ \sim \Phi (t')$. This gives Eqs.~(\ref{final2}-\ref{final3}). 
If we set $\lambda =0$ in Eq.~(\ref{kpz2}) (assuming the scaling regime of constant $v(t)$), 
then the resulting EW equation is invariant under the same transformations providing $z=2$, 
$2\alpha +1=z$ and $\nu \to \nu'$, $\nu' p^2 = \nu$. This, again, gives 
Eqs.~(\ref{final2}-\ref{final3}) but with $z=2$ and $\alpha =1/2$. Transformations 
defined by Eq.~(\ref{trans1}) can be seen as superpositions, where the first scaling
\begin{equation}
\label{trans2}
x \to x/L \, , \;\; h \to h/L^\alpha \, , \;\;  t \to t/L^z  
\end{equation}
is followed by 
\begin{equation}
\label{trans3}
x \to x \, , \;\; h \to p h \, , \; \; t \to p^2 t \, .
\end{equation}
(The order in which transformations (\ref{trans2}) and (\ref{trans3}) are 
superimposed to form transformation (\ref{trans1}) can be reversed.) 
The second transformation, defined by Eq.~(\ref{trans3}), leaves the KPZ equation 
(\ref{kpz2}) invariant provided that new coefficients $\nu'$ and $\lambda'$ 
are related to the old coefficients 
by the relations $\nu' p^2 =\nu$ and $\lambda' p^3= \lambda$ (which is the result 
obtained before) and the old velocity $v(t)$ changes to a new velocity 
$v'(t)=v(t')/p$. Notice, in the absence of the RD admixture we have $p=1$ and 
the transformation (\ref{trans3}) is the identity. Then transformation 
(\ref{trans1}) is the usual FV scaling, 
Eqs.~(\ref{Family1}-\ref{Family2}), with proper choices for $\alpha$ and $z$, 
depending on the process at hand (either KPZ or EW). When the RD process is 
present we have $p<1$. Then, inverting transformation (\ref{trans3}) gives the 
changes in the local-height field $h/p$ and in time scales $t/p^2$. Since the 
elongation of the time scale is inversely proportional to $p^2$ this effect is 
more pronounced than the stretching in $h$. These phenomena take place for all 
$t \ge0$ since transformation (\ref{trans3}) is valid for all $t$, while 
the FV scaling or, equivalently, affine mapping given by Eq.~(\ref{trans2}) 
takes place only in steady-state simulations at times $t \gg t_0(p)$ when the 
system has lost the memory of the initial condition.

One possibility for the behavior seen in this paper for an admixture of the 
KPZ fixed point with the RD fixed point could be for a reason similar to the 
floating-fixed point seen in critical phenomena \cite{FFS01,FFS02,FFS03}. In that 
case two physical fixed points (in integer dimensions) are joined by a line of fixed 
points that are inaccessible (in that case in non-integer dimensions). The result is 
that for a finite system size, and depending on the boundary conditions involved, 
there is an effective critical exponent \cite{FFM01,FFM02,FFM03,FFM04}.  
Only in the limit of infinitely large systems is the physical 
fixed point approached. The effective critical exponents along this 
line of fixed points satisfy (hyper)scaling relationships, just as here 
the relationship $\alpha+z=2$ holds. Further investigations studying much larger 
systems would be needed to see if the floating-fixed point picture holds in our case 
where properties of both the RD and KPZ fixed points are mixed into the  
nonequilibrium surface model.

In summary, the only effect of the RD admixture to either a genuine KPZ or 
EW process is the dilatation of growth scales. The RD blending does not change 
the universality class of the interface since it does not change the dynamics of 
mechanisms that are responsible for building correlations. However, such dilatation, 
when combined with the initial flat-substrate condition, may obscure a clear observation 
of KPZ scaling in simulations as well as in experiments.

\subsection{Roughness \label{roughness}}

\begin{figure}[tp]
\includegraphics[width=7.0cm]{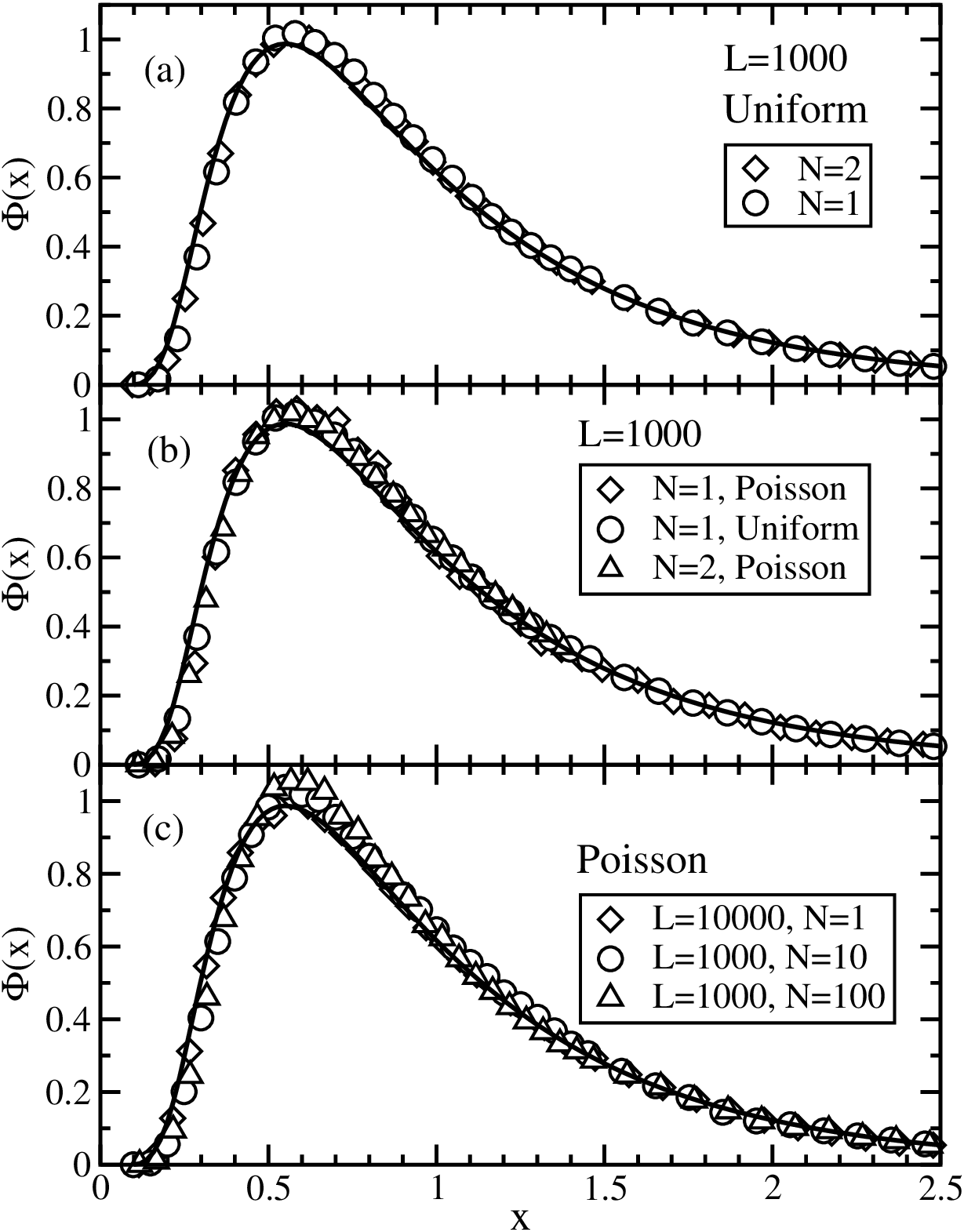}
\caption{\label{pre04f14}
Distributions $\Phi(x)$ of the interface widths at saturation,
$x=w^2/\langle w^2 \rangle$. Results of simulations (symbols) are
compared to Eq.~(\ref{Fotlin}) (continuous curve).
(a) Uniform deposition when $L=1000$ and $N=1,2$ (binning interval $\Delta_U = 0.05$;
the data for $N=1$ fall on the top of data for $N=2$).
(b) Uniform and Poisson depositions when  $L=1000$ and $N=1,2$ ($\Delta_U = 0.05$,
$\Delta_P = 0.5$). (c) Poisson deposition when $L=1000, 10000$ and
$N=1, 10, 100$ ($\Delta_P = 1$).
}
\end{figure}

In this section we discuss finite-size effects in scaling for the VTH interfaces 
simulated by the worst-case-scenario deposition/update rules $1$ and $2$. Despite that 
these interfaces belong to the KPZ universality class, the precise value of the 
roughness exponent depends on the lattice size $L$ and on the type of deposition. 
In the case of Poisson deposition it requires large $L$ to attain scaling with 
$2\alpha \approx 1$ \cite{KTN+00} ($\alpha$ approaches $1/2$ from below as $L$ is 
increased), while for uniform depositions the required $L$ is approximately two 
orders smaller than that for Poisson depositions.

At saturation the growth of a (1+1) dimensional KPZ (or EW) surface can be mapped 
onto a diffusion problem with column-height fluctuations 
$\delta h \sim 1/z_w$, where $z_w$ is the dynamic exponent of a random 
walker \cite{BS95,Odor04} that connects with the roughness exponent, 
$\alpha z_w =1$. The exact $\alpha = 1/2$ indicates the total lack of 
correlations and $\alpha > 1/2$ indicates their presence. The random-walk 
interfaces are characterized by the following width distribution function 
$\Phi (w^2/ \langle w^2 \rangle)$ \cite{FOR+94}
\begin{equation}
\label{Fotlin}
\Phi (x) = \frac{\pi ^2}{3} \sum_{n=1}^{\infty} (-1)^{n-1} n^2 
\exp \left( -\frac{\pi ^2}{6} n^2 x \right) .
\end{equation}
In simulations of growth models where $w^2$ depends on a single length scale, 
$\Phi$ is obtained by normalizing a histogram $P(w^2)$ of the width distribution 
\cite{FOR+94,RP94,PRZ94,AR96,MPP+02}
\begin{equation}
\label{Racz}
\Phi \left( \frac{w^2}{\langle w^2 \rangle} \right) = \langle w^2 \rangle P (w^2) . 
\end{equation}
This technique of describing surfaces in terms of their scaling functions 
$\Phi$ gives very good agreement between theoretical functions $\Phi$ 
(whenever available) and simulated curves for growth models with integer 
height increments \cite{FOR+94,RP94,PRZ94,AR96,MPP+02}. Figure~\ref{pre04f14} 
shows $\Phi$ obtained in simulations with several $N$ and $L$ for our 
deposition/update models with Poisson and uniform depositions. We observe that 
these curves closely follow the theoretical curve given by Eq.~(\ref{Fotlin}). 
In the computation of the quantities in Eq.~(\ref{Racz}) (see Appendix) we used 
a variable step size $\Delta$ in binning the $w^2$ data ($\Delta=0.05, 0.1, 0.5$ and $1$) 
and a variable number $N_{data}$ of data points at saturation 
($6.4 \times 10^6 < N_{data} < 10^8$) to ensure that the results of 
Fig.~\ref{pre04f14} represent the true limit of $\Phi$ obtained in our models. 

The exact collapse of the distributions $\Phi$ obtained in simulations on the 
theoretical curve of Fotlin {\it et al.} \cite{FOR+94}, given by Eq.~(\ref{Fotlin}), 
is not related to the type of the deposition, as is illustrated by the data 
in Fig.~\ref{pre04f14} and is observed even for small system sizes ($L \propto 100$). 
Therefore explaination for the sensitivity of $\alpha$ to $L$ and to the deposition type 
must lie elsewhere. For moderate to 
large $L$ presented in this work, when $N=1$ data collapse at saturation required 
$2\alpha \cong 0.94$ for uniform deposition, $2\alpha \cong 0.92$ for Gaussian 
deposition, and $2\alpha \cong 0.88$ for Poisson deposition. Similarly, when $N=2$ 
the collapse was achieved with $2\alpha \cong0.94$ for uniform deposition and 
$2\alpha \cong 0.9$ for Poisson deposition. This variation of $\alpha$ with the 
deposition type suggest that a small departure $\delta _{\alpha} = 1-2\alpha$ from 
the exact value $2\alpha=1$ is a nonuniversal parameter. As observed, the largest 
difference $\delta _{\alpha}$ is for depositions that have the largest variance 
(i.e., Poisson) and the smallest $\delta _{\alpha}$ is for depositions of the smallest 
variance (i.e., uniform). This observation strongly indicates that $\delta _{\alpha}$ 
is  related to the system memory (introduced in Sec.~\ref{flat}), i.e., to time scales $T$ 
on which the interface does not remember past depositions. In other words, 
$\delta _{\alpha}$ depends on the minimal interval $T$ such that, on the average, 
a deposition at time $t$ does not affect  depositions at time $(t+T)$. Long memory 
scales $T$ lead to a build-up of temporal correlations, thus, produce a variation to 
Gaussian noise in Eq.~(\ref{kpz}), possibly modifying noise strength $D$ in 
Eq.~(\ref{noise}). Such a variation may influence the way in which the system attains 
the stationary state \cite{Fog01} and, possibly, the value of $\alpha$. We leave this issue 
open to future investigations.

\subsection{False scaling \label{false}}

\begin{figure}[tp]
\includegraphics[width=8.5cm]{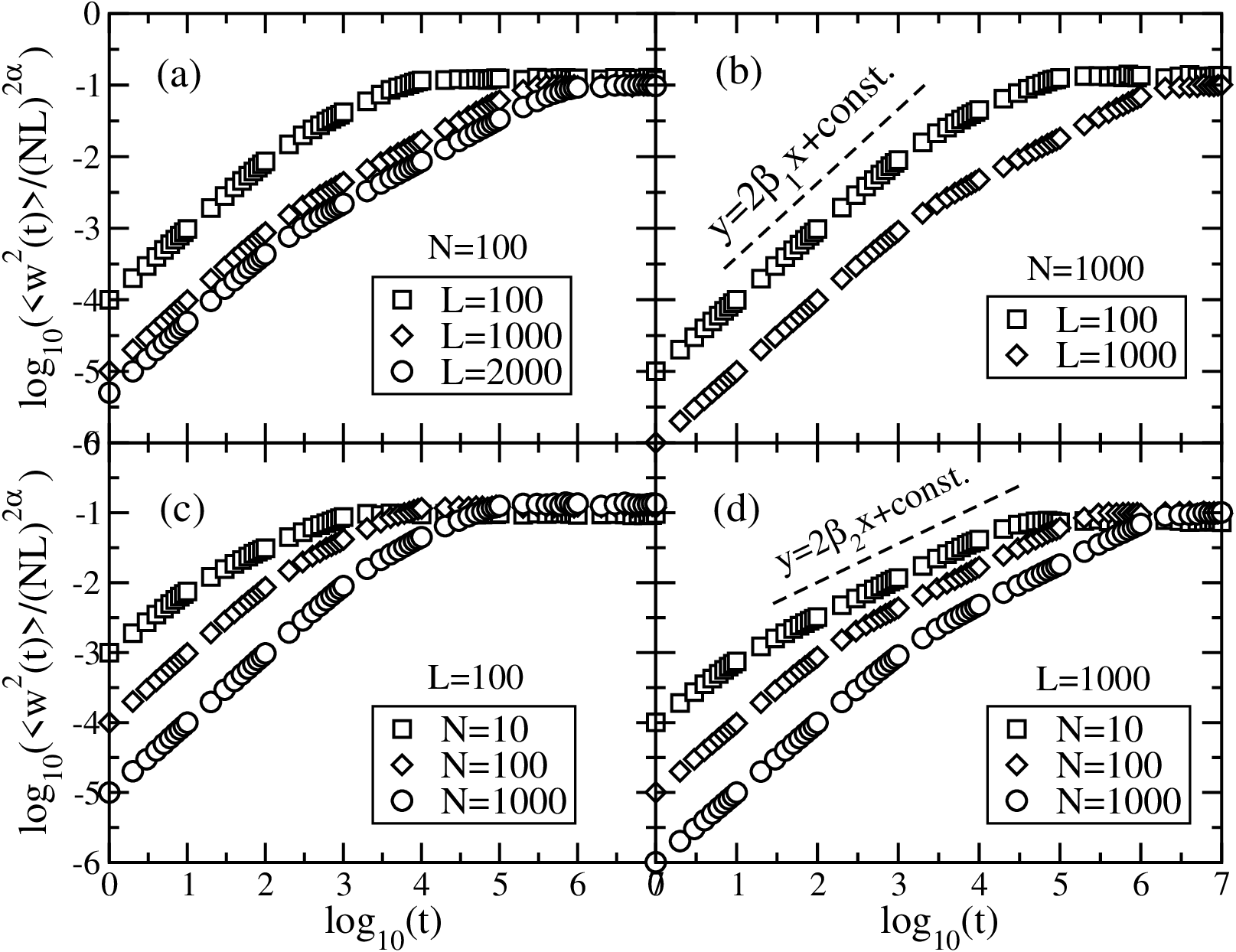}
\caption{\label{pre04f15}
The false-scaling time evolution for $t>0$ of the scaled widths, obtained with
Poisson depositions in rule $3$: (a) $N=100$; (b) $N=1000$; (c) $L=1000$; (d) $L=1000$.
}
\end{figure}

\begin{figure}[tp]
\includegraphics[width=7.0cm]{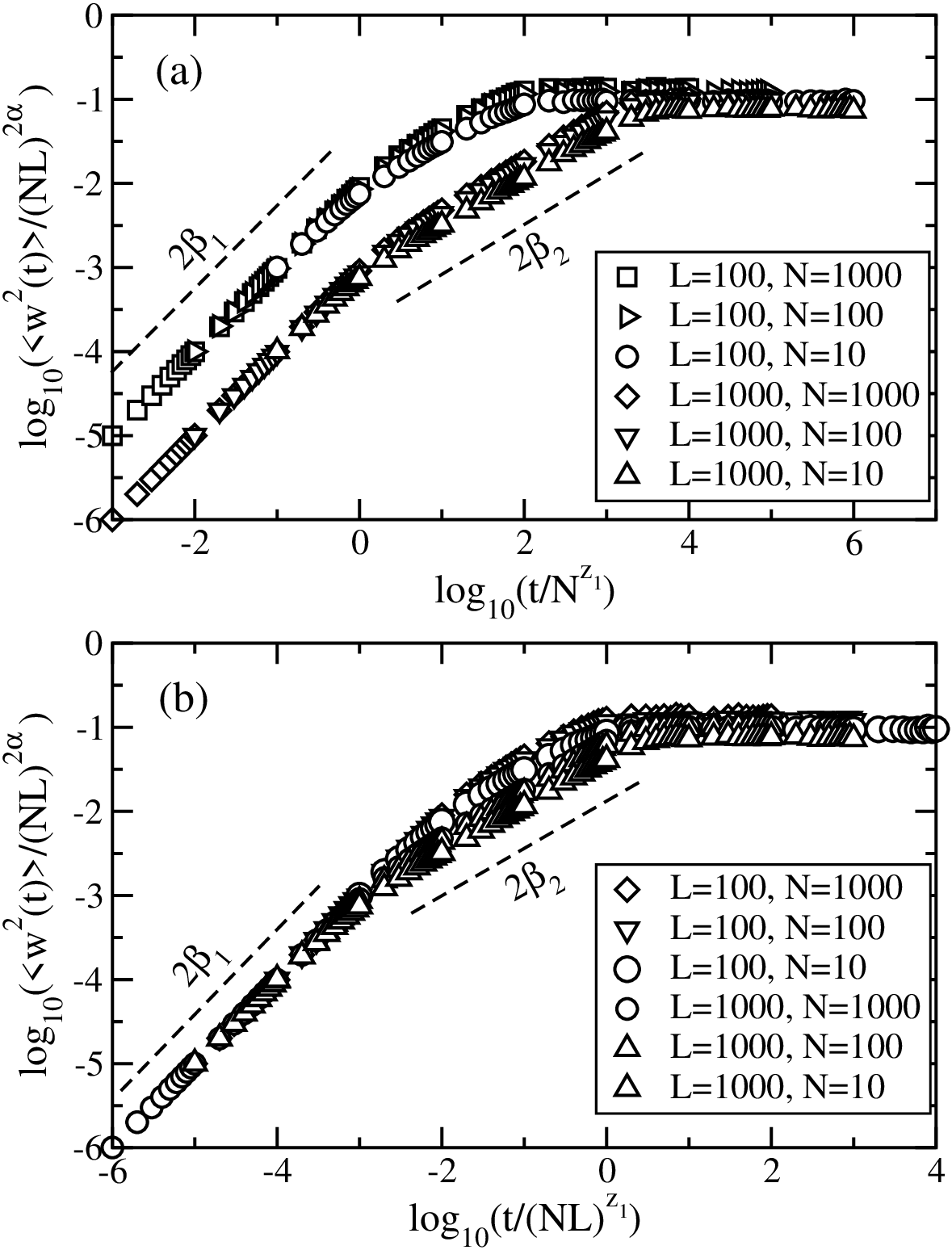}
\caption{\label{pre04f16}
False scaling for the curves of Fig.~\ref{pre04f15}.
(a) Scaling with $N$ in  $t$, $z_1=1$.
(b) Scaling with $L$ in $t$ for the data in figure (a).
}
\end{figure}

A blind application of the scaling technique to numerical data sets, 
without supporting analysis of temporal scales and invariants involved, 
may lead to false conclusions. For example, assuming a global linear behavior of 
$\langle w^2 \rangle$ vs the variable $(NL)$ (Fig.~\ref{pre04f11}{\it c}), it is 
possible to collapse the saturated widths by using an effective exponent 
$2\alpha \approx 1$. Performed for all $t>0$, such scaling shows two 
distinct growth regimes in the evolution curves (Fig.~\ref{pre04f15}). 
For each $L$, these can be further collapsed into groups by scaling 
$t \to t/N^{z_1}$ for all $t$, taking $z_1=1$ (Fig.~\ref{pre04f16}{\it a}). 
The excellent total data collapse in the first growth regime (of slope 
$2\beta _1$ in Figs.~\ref{pre04f15}-\ref{pre04f16}) is obtained when 
$t \to t/(NL)^{z_1}$, but at the expense of only an approximate collapse in 
the second growth regime (of slope $2\beta _2$ in Figs.~\ref{pre04f15}-\ref{pre04f16}). 
This final result (Fig.~\ref{pre04f16}{\it b}) seems to look plausible because 
$2\beta_1=1$ and $2\alpha=1$ give a  hypothetical exponent $z_1=1$, and the curves 
fall exactly on the top of each other in the early growth phase and at saturation. 
However, upon a closer inspection it appears that the length of the intermediate 
regime of slope $2\beta_2$ expands when $L$ is increased. In the limit of large but 
finite $L$, the scaled curves in Fig.~\ref{pre04f16}{\it b} will cover the lower right-hand 
semi-plane, bounded by lines $y= \log_{10}(\langle w^2 \rangle /(NL)^{2\alpha}) = \mathrm{const.}$ 
and $y= 2\beta_1 \log_{10}(t/(NL)^{z_1}) + \mathrm{const}$. Hence, there is no scaling. 
In fact, this example supports our earlier conclusion that the initial growth phase 
for $t<t_0(N)$, the artifact of the initial condition, does not scale.

\subsection{Application to PDES \label{application}}

In modeling of the conservative update mode in PDES, we represent sequential events on 
a processor in terms of their corresponding local virtual times. The column height that 
rises at the $k$th lattice site in the simulated VTH represents the total time of 
operations performed by the $k$th processor. These operations can be seen as a sequence 
of update cycles, where each cycle has two phases. The first phase is the processing of 
the assigned set of discrete events (e.g., spin flipping on the assigned sublattice). 
This phase is followed by a messaging phase that closes the cycle, when a processor 
broadcasts its findings to other processors. But the messages broadcasted by other 
processors may arrive any time during the cycle. Processing related to these messages 
(e.g., memory allocations/deallocations, sorting and/or other related operations) 
are handled by other algorithms that carry their own virtual times. In fact, 
in actual simulations, this messaging phase may take an enormous amount of time, depending 
on the hardware configuration and the message processing algorithms. In our modeling the 
time extent of the messaging phase is ignored as though communications among processors 
were taking place instantaneously. In this sense we model an ideal system of processors. 
The local virtual time of a cycle represents only the time that logical processes require 
to complete the first phase of a cycle. Therefore, the spread in local virtual times 
represents only the desynchronisation that arises due to the conservative algorithm alone.

The measure of this desynchronization is provided by $\langle w(t) \rangle$. 
Since in PDES the memory request per processor, required for past-state savings, 
is determined by the extent to which processors get desynchronized, 
$\langle w(t) \rangle$ can be considered as an indirect approximate measure of 
this memory request. Its growth during the entire time span of the PDES 
computations is given explicitly by Eq.~(\ref{evolution}).

We showed that in conservative PDES, given the PDES size (finite $L$ and $N$),  
this memory request does not grow without limit 
but varies as the computations evolve. The fastest growth, proportional to $\sqrt{t}$, 
characterizes the initial start-up phase. The length of the start-up phase depends on 
the load per processor (represented here by $N$). The start-up phase is characterized by 
decreasing values of both the utilization and the progress of the global simulated time 
(i.e., the smallest local virtual time from all processors at each simulation step). 
In the steady-state simulations, when the utilization has already stabilized at a mean 
constant value (and so has the progress of the global simulated time), the memory request 
grows slower, at a decreasing rate $\sim 1/t^{2/3}$. In this phase, the mean request can 
be estimated globally from Eq.~(\ref{final}) or Eq.~(\ref{explicit}). The important 
consequence of scaling, expressed by Eq.~(\ref{final}), is the existence of the upper 
bound for the memory request for any finite number $L$ of processors and for any finite 
load $N$ per processor. As it is stated by Eq.~(\ref{explicit}), on the average, this 
upper bound increases proportionally to $\sqrt{NL}$ with the size of conservative PDES. 

The characteristic time scale $t_0(N)$ from the first step to the steady-state simulations 
can be estimated by monitoring the utilization for the minimal processor load 
(to determine $t_0$)  and, subsequently, scaling this time with $N$. Similarly, 
the characteristic time scale to $t_\times(N)$, when the desynchronization reaches 
its steady state, can be scaled with the processor load to determine an approximate 
number of simulation steps to the point when the mean memory request does not grow anymore.

During the steady state simulations the utilization $\langle u \rangle$ is given by 
the approximate Eq.~(\ref{steadyu}), where $p(N)=\sqrt{2/N}$. The smallest utilization 
$\langle u_p \rangle$ for any processor load is obtained by taking the limit 
$L \to \infty$ of Eq.~(\ref{steadyu}):
\begin{equation}
\label{minimau}
\langle u_p \rangle = (1-p(N)/2)(1-p(N)/4) .
\end{equation}
Since as $N \to \infty$, $p(N)$ decreases as $\sim 1/\sqrt{N}$, Eq.~(\ref{minimau}) 
shows that $\langle u_p \rangle$ grows very fast when the processors' load increases. 
For example, $\langle u_p \rangle$ for $N=100$ is about $90 \%$, close to its asymptotic 
limit of $100 \%$. For the minimal processors' load $(p=1)$ 
Eq.~(\ref{steadyu}) gives for $L \ge 3$
\begin{equation}
\label{minu}
\langle u_0 \rangle = \frac{3}{8} + \frac{1}{8L} .
\end{equation}
In the limit $L \to \infty$, $\langle u_0 \rangle$ is the smallest possible value of 
the utilization. As Eq.~(\ref{minu}) shows, this value is a non-zero constant 
(equal to $3/8$ when derived from simulations with Poissonian distribution of 
waiting times). This nonzero lower bound on the  utilization 
and the finite upper bound for the memory request for finite $L$ show that  
conservative PDES are generally scalable with the number of computing processors 
when performed in the ring communication topology. 
The extension of this conclusion to other communication topologies 
requires a separate study.

\section{SUMMARY AND CONCLUSIONS \label{summary}}

We considered a two-component growth  
in (1+1) dimensions. One of the components is RD  
that takes place with probability $(1-p)$. The other component, which 
takes place with probability $p$, is a deposition process that 
generates correlations typical of  
KPZ dynamics. The growth is simulated from an initially flat substrate.

We show that the flat-substrate initial condition is responsible for the existence 
of the initial non-scaling regime in simulations. The length of this initial 
phase is a nonuniversal parameter (it depends on the type of depositions and on 
the particulars of the model). However, its presence  
is a universal phenomenon.

During the initial phase the simulations relax to a steady state. For the 
models considered in this work, the transition time  
to the steady state can be defined as the time when the mean interface velocity 
attains a constant value. We showed that for these models the mean interface velocity 
is a multiple of the mean utilization.

During the steady state the interface width satisfies FV scaling. 
We derived the universal scaling function for the width and showed that 
the RD admixture acts as a dilatation mechanism to the time and 
height scales, but leaves the KPZ correlations intact. This conclusion has been 
generalized to two-component models that mix RD with depositions that classify 
within the EW universality class. In particular, we showed that the RD admixture 
is responsible for the $p$ dependent affine change of scales ($h \to h/p$ and $t \to
t/p^2$) that is superimposed on the usual scaling and leaves the dynamics invariant.

The models, studied in this work, that give rise to the KPZ 
correlations are the Poisson-, the Gauss-, and the uniform-random depositions 
either to local interface minima or to local minima and randomly 
selected local slopes. Despite that the simulated interfaces 
belong to the KPZ universality class, the precise value of their roughness exponent 
depends on the deposition type. This observation suggests that such noisy deposition 
mechanisms may produce relatively long-scale temporal correlations. Secondly, this 
small departure from the exact value is nonuniversal. Further studies are required  
to investigate this issue.

In application to conservative PDES, we showed that 
the memory request per processor, required for state savings, 
does not grow without limit for a finite number of processors and a finite load per processor 
but varies as the PDES evolve. The important 
consequence of the derived scaling is the existence of the upper bound for 
the desynchronization, thus, for this memory request. Also,  
the utilization of the parallel processing environment has a non-zero lower 
bound as the number of processors increases infinitely. Thus, the conservative 
PDES are generally scalable in the ring communication topology.

\begin{acknowledgments}
The authors thank P. A. Rikvold, R. B. Pandey and G. Korniss for stimulating discussions. 
This work is supported by NSF grants DMR-0113049 and DMR-0120310; and by 
the ERC Center for Computational Sciences at MSU.
This research used resources of the National Energy Research
Scientific Computing Center, which is supported by the Office
of Science of the US Department of Energy under contract No.
DE-AC03-76SF00098.\\
\end{acknowledgments}

\appendix*
\section{Distributions}
\label{histograms}

Both $h_k(t)$ 
and $w^2(t)$ are real numbers that take on continuous values. 
Suppose $S=\{f_1, f_2, ..., f_N\}$ is a set of $N$ such numbers obtained in 
simulations. Let $a=\min(S)$ and $b=\max(S)$. The interval $(b-a)$ is 
partitioned into $M$ segments, each of length $\Delta = (b-a)/M$. 
Each segment is a bin that is indexed by its left end $y_i = a + (i-1) \Delta$, 
$i=1, 2, ..., M$. Let $m_i$ be the multiplicity of the $i$th bin, i.e., 
$m_i$ is the number of points from $S$ that fall between $y_i$ and $y_{i+1}$. 
The mean value of $S$ is
$\langle f \rangle = 1/N \sum_{k=1}^{N} f_k 
= \sum_{i=1}^{M} P_i \, \langle f_i \rangle$, 
where $\langle f_i \rangle = (1/m_i) \sum_{n=1}^{m_i} f_{in}$ is 
the mean value taken on a subset of $S$ that belongs to the $i$th bin, 
and $P_i$ is the frequency function $P_i = (m_i/N)(1/\Delta)$. 
The abscissas $x_i$ and function values $\Phi_i = \Phi(x_i)$ are: 
$x_i= \langle f_i \rangle / \langle f \rangle$ and  
$\Phi_i = \langle f \rangle P_i$.
These values are plotted in Fig.~\ref{pre04f06} and Fig.~\ref{pre04f14}. 
Here, $\Phi (x)$ is properly normalized 
$\int_{0}^{\infty} dx \, \Phi (x) = 1$. 
The absolute spread $(b-a)$ determines a suitable $\Delta$ in the 
computation of $\langle f_i \rangle$ within acceptable precision. This gives 
the total number $M$ of bins. For a given $M$, the accuracy of $P_i$ depends on $N$.

\end{document}